\documentclass[aps,twocolumn]{revtex4}

\RequirePackage[T2A]{fontenc}

\usepackage{amssymb}
\usepackage{amsmath}
\usepackage{bm}
\usepackage[dvips]{epsfig}
\usepackage{graphicx}
\usepackage{textcomp}

\begin{document}

\title{Ultra-complex conductivity diagrams in the nearly free 
electron approximation} 

\author{A.Ya. Maltsev}

\affiliation{
\centerline{\it L.D. Landau Institute for Theoretical Physics}
\centerline{\it 142432 Chernogolovka, pr. Ak. Semenova 1A,
maltsev@itp.ac.ru}}

\begin{abstract}
We investigate the possibility of the emergence of ultra-complex 
conductivity diagrams in the nearly free electron 
approximation for metals with cubic symmetry. Estimates show 
that the emergence of such diagrams requires the Fermi level to 
fall into very narrow energy intervals within the conduction band. 
In our view, this circumstance is mostly due to the high symmetry 
and the simplest analytical form of the dispersion relations 
$\epsilon ({\bf p})$ under consideration. 
\end{abstract}

\maketitle

\section{Introduction}

\begin{figure}[t]
\begin{center}
\includegraphics[width=\linewidth]{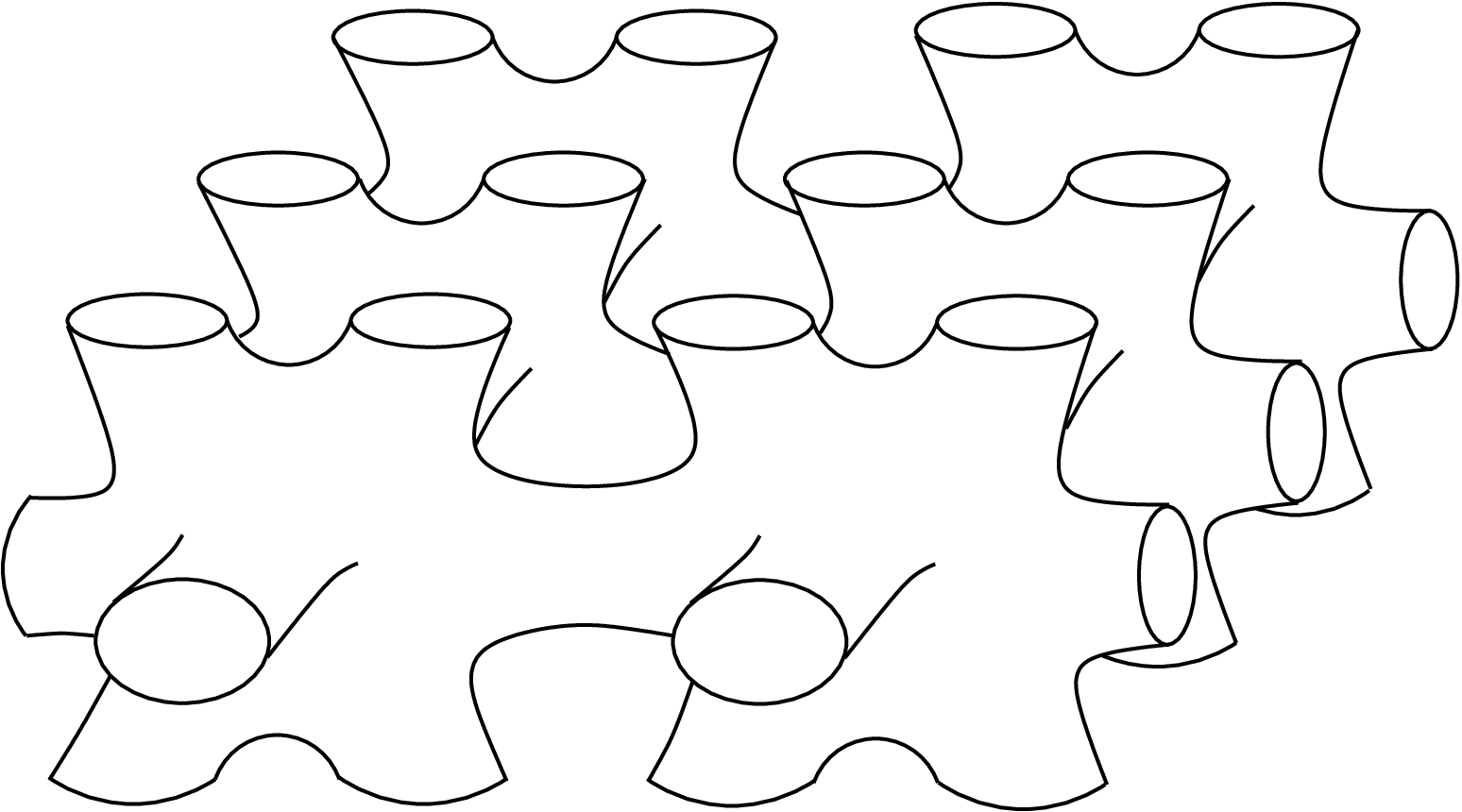}
\end{center}
\caption{Complex Fermi surface in $\, {\bf p}$ - space.}
\label{Fig1}
\end{figure}

 In this paper, we consider the angular diagrams that 
determine the conductivity of normal metals in strong magnetic 
fields. We note that we are interested only in metals with 
sufficiently complex Fermi surfaces extending in all directions 
in $\, {\bf p}$ - space (Fig. \ref{Fig1}).

 As is well known 
(see e.g. \cite{Kittel, Ziman, AshcroftMermin, Abrikosov}), 
the electron quasi-momentum is defined modulo the vectors of the 
reciprocal lattice $\, L^{*} \, $:
\begin{multline*}
{\bf p} \,\,\, \equiv \,\,\, {\bf p} \,\, + \,\, 
n_{1} \, {\bf a}_{1} \,\, + \,\, n_{2} \, {\bf a}_{2} \,\, + \,\,
n_{3} \, {\bf a}_{3} \,\,\, ,  \\  
n_{1} , \, n_{2} , \, n_{3} \,\, \in \,\, \mathbb{Z} \,\,\, , 
\end{multline*}
where the basis of the lattice $\, L^{*} \, $ is given by the vectors
\begin{multline*}
{\bf a}_{1} \,\,\, = \,\,\, 2 \pi \hbar \,\,
{{\bf l}_{2} \, \times \, {\bf l}_{3} \over
({\bf l}_{1}, \, {\bf l}_{2}, \, {\bf l}_{3} )} \,\,\, , \quad \quad
{\bf a}_{2} \,\,\, = \,\,\, 2 \pi \hbar \,\,
{{\bf l}_{3} \, \times \, {\bf l}_{1} \over
({\bf l}_{1}, \, {\bf l}_{2}, \, {\bf l}_{3} )} \,\,\, ,  \\
{\bf a}_{3} \,\,\, = \,\,\, 2 \pi \hbar \,\,
{{\bf l}_{1} \, \times \, {\bf l}_{2} \over
({\bf l}_{1}, \, {\bf l}_{2}, \, {\bf l}_{3} )} \,\,\, , 
\quad \quad \quad \quad
\end{multline*}
and $\, ({\bf l}_{1}, {\bf l}_{2}, {\bf l}_{3}) \, $ 
define the basis of the crystallographic lattice $\, L \, $. 

 As a consequence, the dispersion relation (the dependence of 
energy on quasi-momentum) $\, \epsilon ({\bf p}) \, $ in 
a crystal can be considered either as a 3-periodic function in 
the space $\, \mathbb{R}^{3} \, $ or simply as a smooth 
function on the three-dimensional torus
$$\mathbb{T}^{3} \,\,\, = \,\,\, \mathbb{R}^{3} \Big/ L^{*} $$
determined by the factorization of $\, {\bf p}$ - space by 
the reciprocal lattice vectors. 

 The Fermi surface of a metal
$$S_{F} \, : \quad  \epsilon ({\bf p}) \, = \, \epsilon_{F} $$
can also be viewed either as a 3-periodic surface in the 
$\, {\bf p}$ - space (Fig. \ref{Fig1}) or as a smooth compact 
surface embedded in $\, \mathbb{T}^{3} \, $. We note here that 
both of these representations play very important role when 
considering electronic phenomena in strong magnetic fields.

\begin{figure*}[t]
\begin{center}
\includegraphics[width=\linewidth]{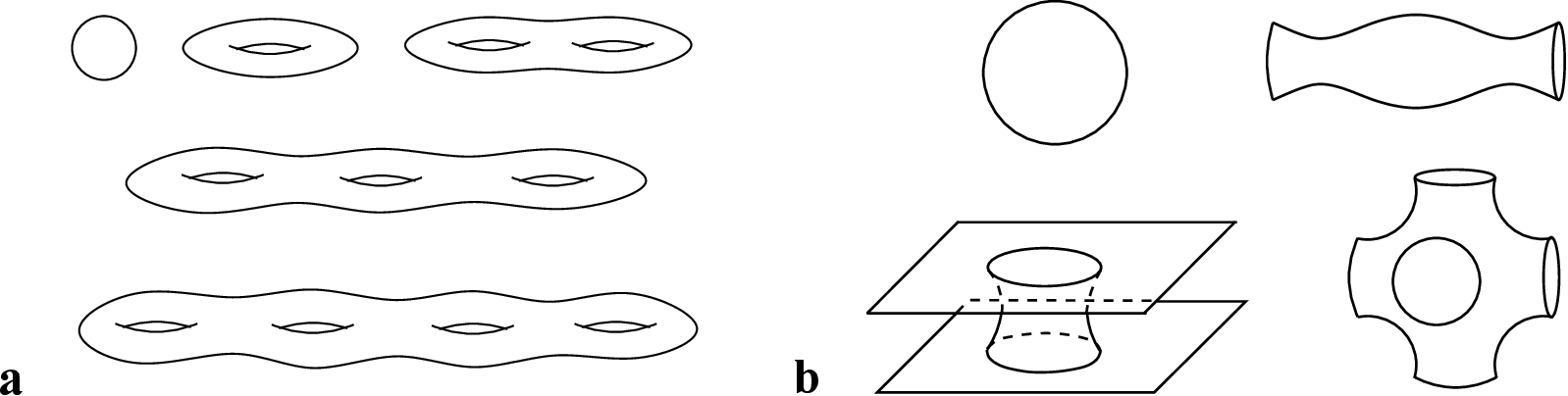}
\end{center}
\caption{(a) Abstract orientable surfaces of genus 
$\, g \, = \, 0, 1, 2, 3, 4 \, $. 
(b) Examples of Fermi surfaces of rank 0, 1, 2 and 3.}
\label{Fig2}
\end{figure*}

 Each compact (orientable) surface $\, S_{F} \, $ has a 
topological genus $\, g \, $, which can take values 
$\, g \, = \, 0, 1, 2, 3, 4, \dots \, $ (Fig. \ref{Fig2}, a).

 The embedding $\, S_{F} \subset \mathbb{R}^{3} \, $ 
(and also $\, S_{F} \subset \mathbb{T}^{3} $) also has 
a topological rank $\, {\rm Rank} \, S_{F} $. By definition, 
the topological rank of $\, S_{F} $ determines the number of 
independent directions of extension of the surface $\, S_{F} $ 
in $\, \mathbb{R}^{3} $ (Fig. \ref{Fig2}, b). It is easy 
to see that the rank of the Fermi surface can take the 
values 0, 1, 2, and 3.

 For topological reasons, the rank of a Fermi surface cannot 
exceed its genus. Thus, Fermi surfaces that are sufficiently 
complex from our point of view have 
$\, {\rm Rank} \, S_{F} \, = \, 3 \, $ and $\, g \geq 3 \, $.

\vspace{1mm}

 The application of an external magnetic field causes an adiabatic 
evolution of the electron states in a crystal, which is described 
by the quasi-momentum evolution determined by the system
\begin{equation}
\label{MFSyst}
{\dot {\bf p}} \,\,\,\, = \,\,\,\, {e \over c} \,\,
\left[ {\bf v}_{\rm gr} ({\bf p}) \times {\bf B} \right]
\,\,\,\, \equiv \,\,\,\, {e \over c} \,\, \left[ \nabla \epsilon ({\bf p})
\times {\bf B} \right] 
\end{equation}

\begin{figure*}[t]
\begin{center}
\includegraphics[width=\linewidth]{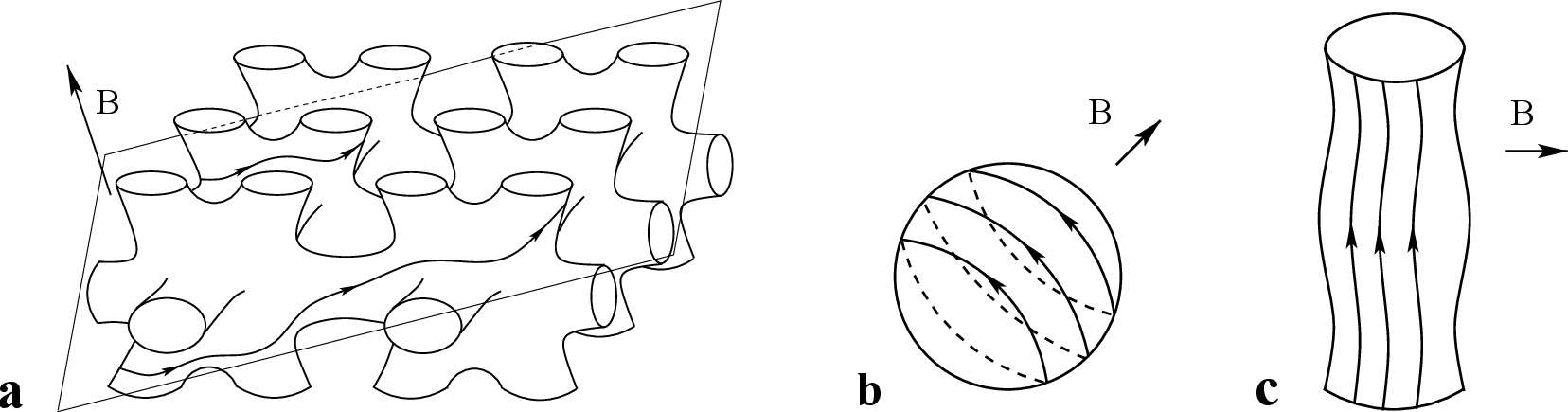}
\end{center}
\caption{(a) Trajectories of the system (\ref{MFSyst}) on the Fermi 
surface of a rather complex shape. (b) Closed trajectories of the 
system (\ref{MFSyst}). (c) Open periodic trajectories of the 
system (\ref{MFSyst}).}
\label{Fig3}
\end{figure*}

 Geometrically, the trajectories of system (\ref{MFSyst}) 
are defined by the intersections of constant-energy surfaces
$$\epsilon ({\bf p}) \,\,\, = \,\,\, {\rm const} $$
and planes orthogonal to $\, {\bf B} \, $. Electronic phenomena 
in metals are determined by trajectories of (\ref{MFSyst}) 
lying on the Fermi surface $\, \epsilon ({\bf p}) = \epsilon_{F} \, $. 
As can be seen, for complex Fermi surfaces, the shape of the 
trajectories of (\ref{MFSyst}) can strongly depend on the direction 
of $\, {\bf B} \, $ (Fig. \ref{Fig3}, a).

 In the limit of large mean free paths and strong magnetic fields, 
electron transport phenomena in a crystal are largely determined 
by the geometry of the trajectories of system (\ref{MFSyst}) in the 
$\, {\bf p}$ - space. In particular, the behavior of the conductivity 
tensor $\, \sigma^{kl} ({\bf B}) \, $ depends significantly on the 
presence or absence of open (unclosed) trajectories of system 
(\ref{MFSyst}) on the Fermi surface (\cite{lifazkag,lifpes1,lifpes2,etm}). 
The corresponding limit can be formally written as the condition 
$\, \omega_{B} \tau \gg 1 \, $, where $\, \omega_{B} \, $ is the 
cyclotron frequency, and $\, \tau \, $ is the electron mean free 
time in the crystal.

 More precisely, it is necessary in fact to require that the electron 
travels a sufficiently large part of its trajectory (compared 
to $\, p_{F}$) between scattering events. It should be noted immediately 
that this limit applies to sufficiently pure single-crystal samples 
at low temperatures ($T \leq 1 {\rm K}$) and strong magnetic 
fields ($B \geq 1 {\rm Tl}$).

\vspace{1mm}

 A striking example of the manifestation of the geometry of 
trajectories of (\ref{MFSyst}) in transport phenomena is the difference 
in the contributions of closed and open periodic trajectories 
(Fig. \ref{Fig3}, b, c) to the conductivity tensor in the limit
$\, \omega_{B} \tau \rightarrow \infty \, $ (\cite{lifazkag})
\begin{equation}
\label{Closed}
\Delta \sigma^{kl}_{\rm closed} \,\,\,\, \simeq \,\,\,\,
{n e^{2} \tau \over m^{*}} \, \left(
\begin{array}{ccc}
( \omega_{B} \tau )^{-2}  &  ( \omega_{B} \tau )^{-1}  &
( \omega_{B} \tau )^{-1}  \cr
( \omega_{B} \tau )^{-1}  &  ( \omega_{B} \tau )^{-2}  &
( \omega_{B} \tau )^{-1}  \cr
( \omega_{B} \tau )^{-1}  &  ( \omega_{B} \tau )^{-1}  &  *
\end{array}  \right) \,\,\, ,   
\end{equation}
\begin{equation}
\label{OpenPeriodic}
\Delta \sigma^{kl}_{\rm periodic} \,\,\,\, \simeq \,\,\,\,
{n e^{2} \tau \over m^{*}} \, \left(
\begin{array}{ccc}
( \omega_{B} \tau )^{-2}  &  ( \omega_{B} \tau )^{-1}  &
( \omega_{B} \tau )^{-1}  \cr
( \omega_{B} \tau )^{-1}  &  *  &  *  \cr
( \omega_{B} \tau )^{-1}  &  *  &  *
\end{array}  \right)  
\end{equation}

 It can be seen that the contribution of closed trajectories to the 
tensor $\, \sigma^{kl} ({\bf B}) \, $ is similar to the contribution 
of the free electron gas, while the contribution (\ref{OpenPeriodic}) 
has a sharp anisotropy in the plane orthogonal to $\, {\bf B} \, $.

 In formulas (\ref{Closed}) - (\ref{OpenPeriodic}), the quantity 
$\, n \, $ plays the role of the charge carrier concentration, and 
$\, m^{*} \, $ is the effective electron mass in the crystal. 
The sign $\, * \, $ here denotes dimensionless constants of order 1.

 Formulas (\ref{Closed}) - (\ref{OpenPeriodic}) describe the 
asymptotic behavior of the tensor $\, \sigma^{kl} (B) \, $ in the 
limit $\, \omega_{B} \tau \rightarrow \infty \, $, in particular, 
each matrix element above is defined up to a constant factor of 
order 1. However, one important remark can be made.

 Namely, if we are talking about the contribution of all closed 
trajectories covering the Fermi surface of the form Fig. \ref{Fig3}b, 
then a more precise expression can be given for the Hall conductivity. 
Thus, in the leading order in $\, 1/B \, $, the quantity 
$\, \sigma^{xy} (B) \, $ has the form
(\cite{Kittel,Ziman,AshcroftMermin,Abrikosov,etm})
\begin{equation}
\label{HallCond}
\sigma^{xy} \,\,\, = \,\,\, - \sigma^{yx} \,\,\, = \,\,\, \pm \,\, 
{e c \over B} \,\, {2 V \over (2 \pi \hbar)^{3}} \,\,\, ,  
\end{equation}
where $\, V \, $ is the volume bounded by the surface $\, S_{F} \, $ 
in the $\, {\bf p}$ - space.
 
 As can be seen, the value (\ref{HallCond}) does not depend on the 
direction of the magnetic field for a given value of $\, B \, $. 
The signs $\, + \, $ and $\, - \, $ in formula (\ref{HallCond}) 
correspond to the electron- and hole-type Fermi surfaces, respectively. 
Recall that for a dispersion relation $\, \epsilon ({\bf p}) \, $, 
such that
$$\epsilon_{\min} \,\,\, \leq \,\,\, \epsilon ({\bf p})
\,\,\, \leq \,\,\, \epsilon_{\max} \,\,\, , $$
the electron-type Fermi surfaces Fig. \ref{Fig3}b arise at values
$\, \epsilon_{F} \, $ close to $\, \epsilon_{\min} \, $, and 
the hole-type Fermi surfaces arise at $\, \epsilon_{F} \, $ close 
to $\, \epsilon_{\max} \, $.

 Note also that for Fermi surfaces containing open trajectories of 
system (\ref{MFSyst}), the relation (\ref{HallCond}) is not satisfied.

\vspace{1mm}

 The problem of describing all possible trajectories of system 
(\ref{MFSyst}) was posed by S.P. Novikov in his work 
\cite{MultValAnMorseTheory} and was then intensively studied in his 
topological school 
(\cite{zorich1,dynn1992,Tsarev,dynn1,zorich2,DynnBuDA,dynn2,dynn3}). 
As a result of the research, a complete classification of all 
trajectories of systems (\ref{MFSyst}) with arbitrary laws 
$\, \epsilon ({\bf p}) \, $, as well as a detailed description of 
their geometric properties, was carried out. In addition, a description 
of all possible modes of conductivity behavior in the limit 
$\, \omega_{B} \tau \rightarrow \infty \, $, 
corresponding to open trajectories of various types, 
has also been obtained by now 
(see, for example, \cite{BullBrazMathSoc,StatPhys,UMNObzor}).

 For a metal with a given Fermi surface $\, S_{F} \, $, it is 
natural to introduce the ``angular diagram'' defined by the system 
(\ref{MFSyst}). Namely, for each direction of $\, {\bf B} \, $:
$${\bf n} \,\,\, = \,\,\, {\bf B} / B \,\,\, \in \,\,\, 
\mathbb{S}^{2} \,\,\, , $$
the angular diagram indicates the type of open trajectories 
(or their absence) on the surface $\, S_{F} \, $. Since each type 
of open trajectories on $\, S_{F} \, $ is associated with a certain 
type of behavior of the tensor $\, \sigma^{kl} (B) \, $ as 
$\, \omega_{B} \tau \rightarrow \infty \, $, we will call such 
diagrams diagrams of conductivity in strong magnetic fields.

\vspace{1mm}

 As is easy to see, for values $\, \epsilon_{F} \, $ close to
$\, \epsilon_{\min} \, $ or $\, \epsilon_{\max} \, $, the conductivity 
diagrams are very simple. Namely, all directions 
$\, {\bf n} \in \mathbb{S}^{2} \, $ correspond here to the presence 
of only closed trajectories on the Fermi surface (Fig. \ref{Fig3}, b). 
The Hall conductivity here is given by the formula (\ref{HallCond}) 
and is of the electron type at 
$\, \epsilon_{F} \rightarrow \epsilon_{\min} \, $ and of the hole type 
at $\, \epsilon_{F} \rightarrow \epsilon_{\max} \, $.

 We will also classify as simple diagrams those that allow only periodic 
open trajectories of (\ref{MFSyst}) on the Fermi surface. As a rule,
such diagrams also arise for values of $\, \epsilon_{F} \, $ that 
are ``not too'' distant from $\, \epsilon_{\min} \, $ or 
$\, \epsilon_{\max} \, $.

\vspace{1mm}

 The main type of open trajectories of (\ref{MFSyst}) are stable open 
trajectories, i.e. trajectories that are stable with respect to all small 
rotations of $\, {\bf B} \, $, as well as small variations in the value 
of $\, \epsilon_{F} \, $.

\begin{figure*}[t]
\begin{center}
\includegraphics[width=0.9\linewidth]{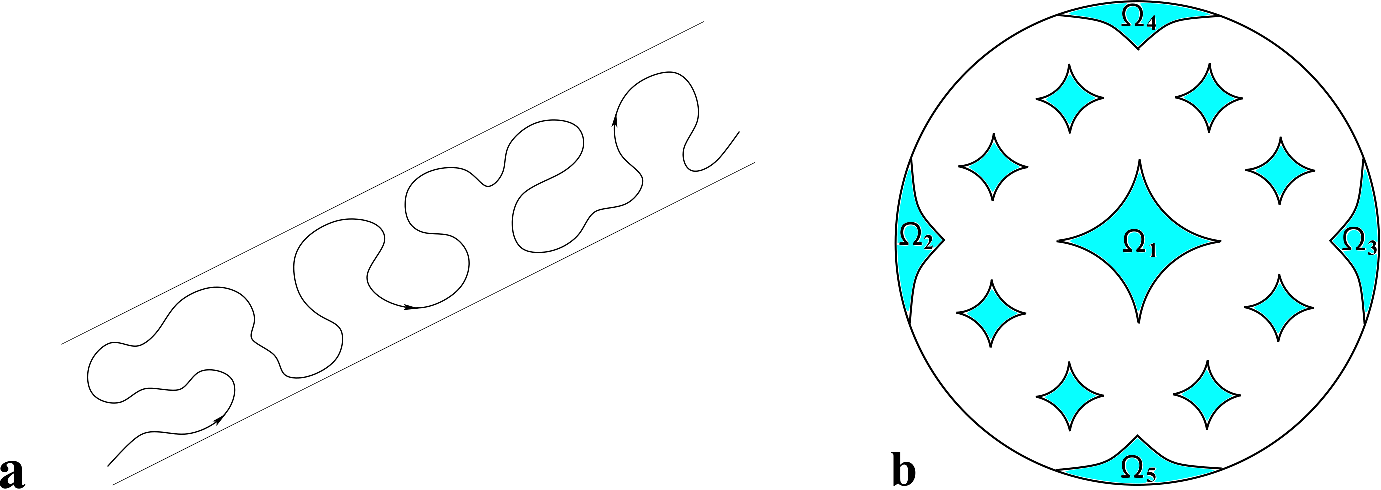}
\end{center}
\caption{(a) Stable open trajectory of system (\ref{MFSyst}) 
in a plane orthogonal to $\, {\bf B} \, $. (b) Stability Zones 
corresponding to different integer planes $\, \Gamma_{\alpha} \, $
on the conductivity diagram (schematically).}
\label{Fig4}
\end{figure*}

 The stable open trajectories of system (\ref{MFSyst}) have 
remarkable geometric properties (\cite{zorich1,dynn1992,dynn1}). 
Namely:

\vspace{1mm}

\noindent
1) Each stable open trajectory of system (\ref{MFSyst}) lies 
(in a plane orthogonal to $\, {\bf B}$) in a straight strip of 
finite width, passing through it (Fig. \ref{Fig4}, a).

\vspace{1mm}

\noindent
2) The mean direction of stable open trajectories 
(in all planes orthogonal to $\, {\bf B}$) for a given 
direction of $\, {\bf B} \, $ is given by the intersection of 
the plane orthogonal to $\, {\bf B} \, $ and some (``integer'') 
plane $\, \Gamma \, $ generated by two reciprocal lattice vectors.

\vspace{1mm}

\noindent
3) The plane $\, \Gamma \, $ is unchanged for small rotations 
of $\, {\bf B} \, $ and variations in the value of 
$\, \epsilon_{F} \, $.

\vspace{1mm}

  Note that, in general, stable open trajectories are not periodic.

\vspace{1mm}

 The emergence of stable open trajectories on the Fermi surface 
corresponds to domains $\, \Omega_{\alpha} \, $ (Stability Zones) 
on the angular diagram, each of which corresponds to its own integer 
plane $\, \Gamma_{\alpha} \, $ (Fig. \ref{Fig4}, b). The Zones 
$\, \Omega_{\alpha} \, $, as well as the integer planes 
$\, \Gamma_{\alpha} \, $, are experimentally observable, 
which formed the basis for the introduction of topological numbers 
in the conductivity of normal metals in \cite{PismaZhETF}.

\vspace{1mm}

 Stable open trajectories can only appear on sufficiently 
complex Fermi surfaces (rank 2 or 3). We will classify an angular 
diagram as sufficiently complex if it contains Stability Zones 
$\, \Omega_{\alpha} \, $.

\vspace{1mm}

 For a fixed dispersion relation $\, \epsilon ({\bf p}) \, $, 
complex angular diagrams arise in some fixed energy range
$$\epsilon_{F} \, \in \, \left( \epsilon^{\cal A}_{1} , \, 
\epsilon^{\cal A}_{2} \right) \,\,\, ,  \quad  
\epsilon_{\min} \,\,\, \leq \,\,\, \epsilon^{\cal A}_{1}
\,\,\, < \,\,\, \epsilon^{\cal A}_{2} \,\,\, \leq \,\,\,
\epsilon_{\max} \,\,\, , $$
where the Fermi surfaces are sufficiently complex.

 The interval
$\, \left( \epsilon^{\cal A}_{1} , \, \epsilon^{\cal A}_{2} \right) \, $
can be divided into 3 intervals
$$\left( \epsilon^{\cal A}_{1} , \, \epsilon^{\cal A}_{2} \right)
\,\,\, = \,\,\, 
\left( \epsilon^{\cal A}_{1} , \, \epsilon^{\cal B}_{1} \right) \cup
\left[ \epsilon^{\cal B}_{1} , \, \epsilon^{\cal B}_{2} \right] \cup
\left( \epsilon^{\cal B}_{2} , \, \epsilon^{\cal A}_{2} \right) $$
(for generic dispersion relations), corresponding to different structures 
of the angular diagrams (see \cite{SecBound,UltraCompl}). This circumstance 
is connected, among other things, with different behavior of the Hall 
conductivity for different values of $\, \epsilon_{F} \, $.

\begin{figure*}[t]
\begin{center}
\includegraphics[width=\linewidth]{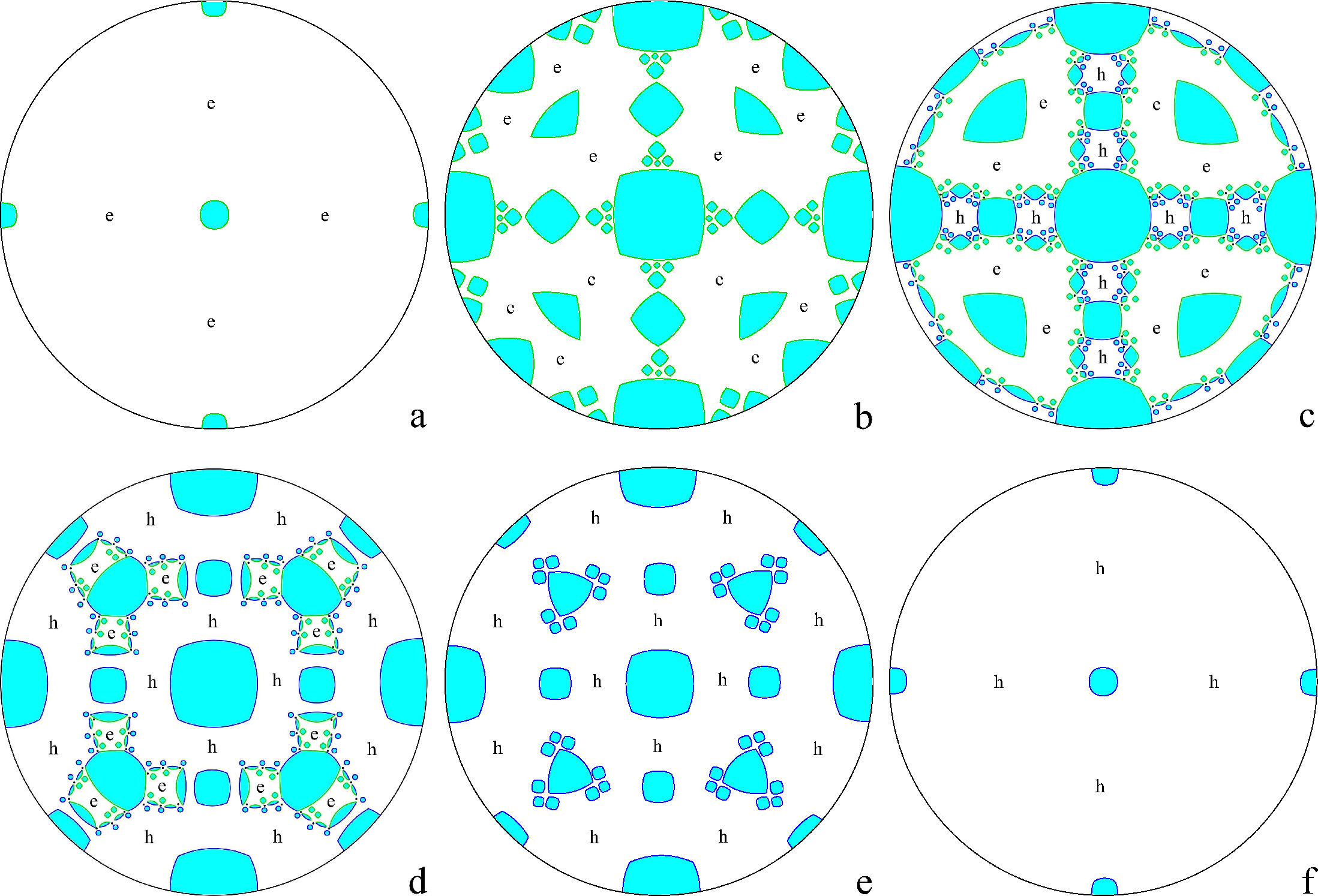}
\end{center}
\caption{Evolution of Stability Zones on the conductivity diagram in 
the interval 
$\, \epsilon_{F} \in \left( \epsilon^{\cal A}_{1} , \, 
\epsilon^{\cal A}_{2} \right) \, $ (schematically). 
The signs e and h mark the regions of electron and hole 
Hall conductivity, respectively.}
\label{Fig5}
\end{figure*}

 Namely, in the evolution of the angular diagrams for 
$\, \epsilon_{F} \in \left( \epsilon^{\cal A}_{1} , \, 
\epsilon^{\cal A}_{2} \right) \, $, 
from the emergence of the first Stability Zones $\, \Omega_{\alpha} \, $ 
to their disappearance (Fig. \ref{Fig5}), one can see that all the diagrams 
contain large regions corresponding to the presence of only closed 
trajectories on the Fermi surface. These regions occupy nearly 
the entire unit sphere for 
$\, \epsilon_{F} \rightarrow \epsilon^{\cal A}_{1} \, $ or 
$\, \epsilon_{F} \rightarrow \epsilon^{\cal A}_{2} \, $ 
(Fig. \ref{Fig5}, a, f) and correspond to electron Hall conductivity 
in the first case, and to hole Hall conductivity in the second case.

 In fact, formulas (\ref{HallCond}) can also be applied 
to complex Fermi surfaces, provided that they contain only closed 
trajectories of the system (\ref{MFSyst}) and this property is 
preserved for (any) small rotations of $\, {\bf B} \, $. 
In this case, the formulas (\ref{HallCond}) take the form
\begin{equation}
\label{HallGen}
\sigma^{xy} \,\,\, = \,\,\, - \sigma^{yx} \,\,\, = \,\,\, \pm \,\, 
{e c \over B} \,\, {2 V_{\mp} \over (2 \pi \hbar)^{3}} \,\,\, ,  
\end{equation}
where $\, V_{-} \, $ is the volume of the region
$$\epsilon ({\bf p}) \,\,\, < \,\,\, \epsilon_{F} \,\,\, , $$
and $\, V_{+} \, $ is the volume of the region
$$\epsilon ({\bf p}) \,\,\, > \,\,\, \epsilon_{F} $$
in the torus
$$\mathbb{T}^{3} \,\,\, = \,\,\, \mathbb{R}^{3} \Big/ L^{*} $$

\begin{figure*}[t]
\begin{center}
\includegraphics[width=\linewidth]{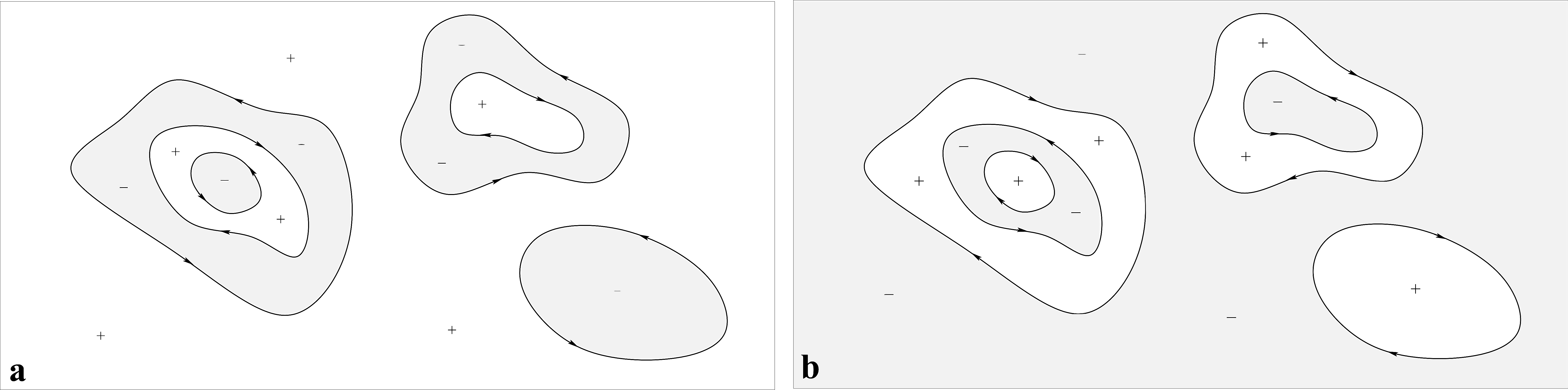}
\end{center}
\caption{Situations $\, A (-) \, $ and $\, A (+) \, $ in planes 
orthogonal to $\, {\bf B} \, $.}
\label{Fig6}
\end{figure*}

 To select the value of $\, V_{-} \, $ or $\, V_{+} \, $ in the 
formula (\ref{HallGen}), it is necessary to consider the picture 
of closed trajectories of (\ref{MFSyst}) in planes orthogonal to 
$\, {\bf B} \, $. Namely, in the described situation, two cases 
are possible:

\vspace{1mm}

\noindent
1) Situation $\, A (-)\, $:

 A plane orthogonal to $\, {\bf B} \, $ contains a unique 
unbounded component of the set
$$\epsilon ({\bf p}) \,\,\, > \,\,\, \epsilon_{F} \,\,\, , $$
and all other components of the sets 
$\, \epsilon ({\bf p}) < \epsilon_{F} \, $ 
and $\, \epsilon ({\bf p}) > \epsilon_{F} \, $ 
are bounded (Fig. \ref{Fig6}, a).

\vspace{1mm}

\noindent
2) Situation $\, A (+)\, $:

 A plane orthogonal to $\, {\bf B} \, $ contains a unique 
unbounded component of the set
$$\epsilon ({\bf p}) \,\,\, < \,\,\, \epsilon_{F} \,\,\, , $$
and all other components of the sets 
$\, \epsilon ({\bf p}) < \epsilon_{F} \, $ 
and $\, \epsilon ({\bf p}) > \epsilon_{F} \, $ 
are bounded (Fig. \ref{Fig6}, b).

\vspace{1mm}

 If the Fermi surface contains only closed trajectories of system 
(\ref{MFSyst}), and this property is stable with respect to small 
rotations of $\, {\bf B} \, $, all planes orthogonal to $\, {\bf B} \, $ 
have the same type ($A (-) \, $ or $\, A (+)$). In formula (\ref{HallGen}), 
the factor $\, V_{-} \, $ corresponds to the situation $\, A (-) \, $, 
and the factor $\, V_{+} \, $ corresponds to the situation $\, A (+) \, $.

\vspace{1mm}

 Near the edges of the interval 
$\, \left( \epsilon^{\cal A}_{1} , \epsilon^{\cal A}_{2} \right) \, $ 
the region
$$\mathbb{S}^{2} \Big\backslash \bigcup_{\alpha} \overline{\Omega_{\alpha}} $$
on the angular diagram is connected, and the number of Zones 
$\, \Omega_{\alpha} \subset \mathbb{S}^{2} \, $ is finite 
(Fig. \ref{Fig5}, a, b, e, f). This type of angular diagrams 
persists in some intervals
$$\epsilon_{F} \in \left( \epsilon^{\cal A}_{1} , \, 
\epsilon^{\cal B}_{1} \right)  \quad \quad \text{and} \quad \quad
\epsilon_{F} \in \left( \epsilon^{\cal B}_{2} , \, 
\epsilon^{\cal A}_{2} \right) \,\,\, , $$
where
$$\epsilon^{\cal A}_{1} \,\,\, < \,\,\, \epsilon^{\cal B}_{1} 
\,\,\, \leq \,\,\, \epsilon^{\cal B}_{2} \,\,\, < \,\,\,
\epsilon^{\cal A}_{2} $$

 Regions where only closed trajectories are present on the Fermi 
surface obviously correspond to electron Hall conductivity
$$\sigma^{xy} \,\,\, = \,\,\, - \sigma^{yx} \,\,\, = \,\,\,  
{e c \over B} \,\, {2 V_{-} \over (2 \pi \hbar)^{3}} $$
in the interval 
$\, \epsilon_{F} \in \left( \epsilon^{\cal A}_{1} , \, 
\epsilon^{\cal B}_{1} \right) \, $ and to hole Hall conductivity
$$\sigma^{xy} \,\,\, = \,\,\, - \sigma^{yx} \,\,\, = \,\,\, - \,\, 
{e c \over B} \,\, {2 V_{+} \over (2 \pi \hbar)^{3}} $$
in the interval
$\, \epsilon_{F} \in \left( \epsilon^{\cal B}_{2} , \, 
\epsilon^{\cal A}_{2} \right) \, $. 

 We will call the conductivity diagrams diagrams of the type 
$\, A_{-} \, $ for 
$\, \epsilon_{F} \in \left( \epsilon^{\cal A}_{1} , \, 
\epsilon^{\cal B}_{1} \right) \, $ and diagrams of the type $\, A_{+} \, $ 
for 
$\, \epsilon_{F} \in \left( \epsilon^{\cal B}_{2} , \, 
\epsilon^{\cal A}_{2} \right) \, $.

\vspace{1mm}

 The interval 
$\, \left[ \epsilon^{\cal B}_{1} , \, \epsilon^{\cal B}_{2} \right] \, $ 
separates the two types of diagrams described above 
($A_{-} \, $ and $\, A_{+}$). For generic dispersion relations 
$\, \epsilon ({\bf p}) \, $, the interval 
$\, \left[ \epsilon^{\cal B}_{1} , \, \epsilon^{\cal B}_{2} \right] \, $ 
has finite width ($\epsilon^{\cal B}_{2} > \epsilon^{\cal B}_{1}$).

 In this paper, we will be interested in the position of the interval 
$\, \left[ \epsilon^{\cal B}_{1} , \, \epsilon^{\cal B}_{2} \right] \, $ 
for the relations $\, \epsilon ({\bf p}) \, $ arising in the nearly free 
electron approximation. 

 For
$\, \epsilon_{F} \in 
\left( \epsilon^{\cal B}_{1} , \, \epsilon^{\cal B}_{2} \right) \, $
the set
$$\mathbb{S}^{2} \Big\backslash \bigcup_{\alpha} \overline{\Omega_{\alpha}} $$
contains regions of both electron and hole Hall conductivity 
(Fig. \ref{Fig5}, c, d). Regions of different Hall conductivity are 
separated by ``chains'' of Zones $\, \Omega_{\alpha} \, $, which in 
generic case contain an infinite number of Zones. We will call 
angular diagrams of this type ultra-complex conductivity diagrams, 
or type B diagrams.

\vspace{1mm}

\begin{figure*}[t]
\begin{center}
\includegraphics[width=\linewidth]{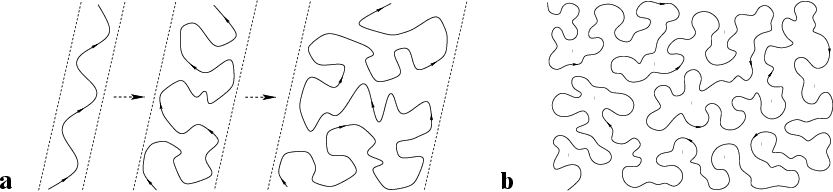}
\end{center}
\caption{(a) Complication of the shape of stable open trajectories 
for small Stability Zones $\, \Omega_{\alpha} \, $. 
(b) Dynnikov's ``Chaotic'' trajectory in a plane orthogonal 
to $\, {\bf B} \, $ (schematically).}
\label{Fig7}
\end{figure*}

 Generic diagrams of type B contain an infinite number of Zones 
$\, \Omega_{\alpha} \, $. Small Zones $\, \Omega_{\alpha} \, $ 
correspond to increasingly complex geometry of the open trajectories 
of system (\ref{MFSyst}) (Fig. \ref{Fig7}, a), which, in turn, 
corresponds to increasingly complex behavior of the tensor 
$\, \sigma^{kl} (B) \, $ in the limit 
$\, \omega_{B} \tau \rightarrow \infty \, $.

 The accumulation points of small Zones $\, \Omega_{\alpha} \, $ 
represent special directions $\, {\bf n} \in \mathbb{S}^{2} \, $, 
corresponding to the emergence of trajectories called ``chaotic''. 
The most common ``chaotic'' trajectories, or Dynnikov's trajectories 
(\cite{DynnBuDA,dynn2}), have a very complex shape, 
wandering ``everywhere'' in planes orthogonal to $\, {\bf B} \, $ 
(Fig. \ref{Fig7}, b). Trajectories of this type give the most complex 
contribution to the tensor $\, \sigma^{kl} (B) \, $, leading to the 
suppression of conductivity along the direction of $\, {\bf B} \, $, 
as well as the emergence of fractional powers of $\, \omega_{B} \tau \, $ 
in the components $\, \sigma^{kl} (B) \, $ (\cite{ZhETF2,TrMIAN}). 
It should be noted that the study of chaotic trajectories of system 
(\ref{MFSyst}) is actively continuing at the present time
(see \cite{zorich3,DeLeo1,DeLeo2,DeLeoPhysLettA,DeLeoPhysB,DeLeo3,
DeLeoDynnikov1,Dynnikov2008,DeLeoDynnikov2,Skripchenko1,Skripchenko2,DynnSkrip1,
DynnSkrip2,AvilaHubSkrip1,AvilaHubSkrip2,DeLeo2017,DynHubSkrip,DynMalNovUMN}).

\vspace{1mm}

 As it turns out, the interval 
$\, \left[ \epsilon^{\cal B}_{1} , \, \epsilon^{\cal B}_{2} \right] \, $ 
is usually quite narrow, and the probability of $\, \epsilon_{F} \, $ 
falling within it is quite low. This explains the difficulty of detecting 
of type B diagrams in real experiments. The paper \cite{TightBind} presents 
estimates for the position of the interval 
$\, \left[ \epsilon^{\cal B}_{1} , \, \epsilon^{\cal B}_{2} \right] \, $ 
and its width in the tight-binding approximation. In this paper, we present 
estimates of the position of this interval in the nearly free electron 
approximation 
(see \cite{Kittel,Ziman,AshcroftMermin,Abrikosov,etm,Harrison}) 
for crystals of cubic symmetry. The presented results can be useful both 
for searching for ultra-complex conductivity diagrams and, possibly, 
for obtaining them by external action on the sample.

\section{Ultra-complex angular diagrams and the case of the 
simple cubic lattice}
\setcounter{equation}{0}

 To determine the location of ultra-complex angular diagrams, it is 
convenient to relate them to the angular diagrams of the entire 
dispersion relation $\, \epsilon ({\bf p}) \, $. The angular diagram 
of a dispersion relation (\cite{dynn3}) specifies the interval 
$\, \left[ \epsilon_{1} ({\bf n}) , \, \epsilon_{2} ({\bf n}) \right] \, $ 
of occurrence of open trajectories of system (\ref{MFSyst}), 
as well as their type, for each direction $\, {\bf B} \, $ 
for a given $\, \epsilon ({\bf p}) \, $. According to \cite{dynn3}:

\vspace{1mm}

\noindent
1) For each direction $\, {\bf n} = {\bf B} / B \, $, open trajectories 
of system (\ref{MFSyst}) arise in a closed interval
$$\epsilon_{F} \,\,\, \in \,\,\, \left[ \epsilon_{1} ({\bf n}) , \, 
\epsilon_{2} ({\bf n}) \right] \,\,\, , $$
which can contract to a point
$$\epsilon_{1} ({\bf n}) \,\,\, = \,\,\, \epsilon_{2} ({\bf n})
\,\,\, = \,\,\, \epsilon_{0} ({\bf n}) $$

\vspace{1mm}

\noindent
2) The case $\, \epsilon_{2} ({\bf n}) > \epsilon_{1} ({\bf n}) \, $ 
corresponds to the emergence of stable or periodic open trajectories 
of (\ref{MFSyst}).

\vspace{1mm}

 For directions $\, {\bf n} \, $ of maximal irrationality, case (2) 
corresponds to the emergence of stable open trajectories of (\ref{MFSyst}) 
that have the same mean direction for all 
$\, \epsilon_{F} \in 
\left[ \epsilon_{1} ({\bf n}) , \, \epsilon_{2} ({\bf n}) \right] \, $. 
The functions $\, \epsilon_{1} ({\bf n}) \, $ and 
$\, \epsilon_{2} ({\bf n}) \, $ are continuous on the set of directions 
$\, {\bf n} \, $ of maximal irrationality and can be extended to continuous 
functions $\, \widetilde{\epsilon}_{1} ({\bf n}) \, $ and 
$\, \widetilde{\epsilon}_{2} ({\bf n}) \, $ on the unit sphere.
The connected components of the set
$$\widetilde{\epsilon}_{2} ({\bf n}) \,\,\, > \,\,\, 
\widetilde{\epsilon}_{1} ({\bf n}) $$
define Stability Zones $\, W_{\alpha} \, $ on $\, \mathbb{S}^{2} \, $, 
corresponding to stable open trajectories associated with different integer 
planes $\, \Gamma_{\alpha} \, $. Stability Zones $\, W_{\alpha} \, $ 
represent domains with piecewise smooth boundaries on the unit 
sphere (\cite{dynn3}).

\begin{figure*}[t]
\begin{center}
\includegraphics[width=0.9\linewidth]{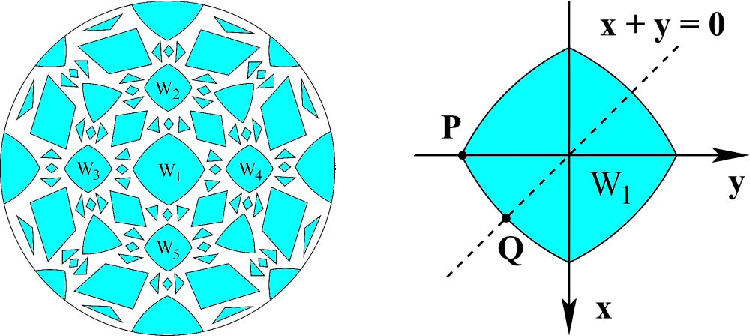}
\end{center}
\caption{Stability Zones $\, W_{\alpha} \, $ on the angular diagram 
for a dispersion relation $\, \epsilon ({\bf p}) \, $ (schematically)
and ``symmetric'' points $P$ and $Q$ on the boundary of Zone $\, W_{1} \, $.}
\label{Fig8}
\end{figure*}

 The Zones $\, W_{\alpha} \, $ form a dense set on the unit sphere 
(Fig. \ref{Fig8}); moreover, for dispersion laws satisfying the condition
$$\epsilon ({\bf p}) \,\,\, = \,\,\, \epsilon (-{\bf p}) \,\,\, , $$
their union is a set of full measure on $\, \mathbb{S}^{2} \, $ 
(I.A. Dynnikov, P. Hubert, P. Mercat, A.S. Skripchenko, in preparation). 
The picture of Zones $\, W_{\alpha} \, $ and behavior of the functions 
$\, \widetilde{\epsilon}_{1} ({\bf n}) \, $,
$\, \widetilde{\epsilon}_{2} ({\bf n}) \, $ 
(as well as $\, \epsilon_{1} ({\bf n}) \, $ and 
$\, \epsilon_{2} ({\bf n}) $) provide important general information 
about a relation $\, \epsilon ({\bf p}) \, $. 

\vspace{1mm}

 Zones $\, \Omega_{\alpha} \, $ (on the conductivity angular diagram) 
represent subdomains of Zones $\, W_{\alpha} \, $, such that
$${\bf n} \, \in \, W_{\alpha} \,\,\, , \quad
\epsilon_{F} \, \in \, \left[ \widetilde{\epsilon}_{1} ({\bf n}) ,
\widetilde{\epsilon}_{2} ({\bf n}) \right] $$
(if this set is not empty).

 As can be seen, for the values of $\, \epsilon^{\cal A}_{1} \, $ 
and $\, \epsilon^{\cal A}_{2} \, $ we have the relations
$$\epsilon^{\cal A}_{1} \,\,\, = \,\,\, \min_{\mathbb{S}^{2}} \, 
\widetilde{\epsilon}_{1} ({\bf n}) \quad , \quad \quad
\epsilon^{\cal A}_{2} \,\,\, = \,\,\, \max_{\mathbb{S}^{2}} \, 
\widetilde{\epsilon}_{2} ({\bf n}) $$

\vspace{1mm}

 For directions of $\, {\bf B} \, $ of maximal irrationality, 
the condition $\, \epsilon_{F} < \widetilde{\epsilon}_{1} ({\bf n}) \, $ 
means falling into the region of the presence of only closed 
trajectories on the Fermi surface (on the conductivity diagram) 
with the Hall conductivity of the electron type. Similarly, 
the condition $\, \epsilon_{F} > \widetilde{\epsilon}_{2} ({\bf n}) \, $ 
for such directions of $\, {\bf B} \, $ means falling into the region of 
the presence of only closed trajectories on the Fermi surface with 
the Hall conductivity of the hole type.

 The presence of regions of both types on the conductivity angular 
diagram means the condition
$$\min_{\mathbb{S}^{2}} \, \widetilde{\epsilon}_{2} ({\bf n}) 
\,\,\, < \,\,\, \epsilon_{F} \,\,\, < \,\,\, 
\max_{\mathbb{S}^{2}} \, \widetilde{\epsilon}_{1} ({\bf n}) \,\,\, , $$
that entails the relations
\begin{equation}
\label{eB12}
\epsilon^{\cal B}_{1} \,\,\, = \,\,\, \min_{\mathbb{S}^{2}} \, 
\widetilde{\epsilon}_{2} ({\bf n}) \quad , \quad \quad
\epsilon^{\cal B}_{2} \,\,\, = \,\,\, \max_{\mathbb{S}^{2}} \, 
\widetilde{\epsilon}_{1} ({\bf n}) \,\,\, , 
\end{equation}
which determine the interval of emergence of type B conductivity 
diagrams (\cite{SecBound,UltraCompl}).

 Let us immediately note that for ``physically realistic'' dispersion 
relations we will assume that all Zones $\, \Omega_{\alpha} \, $ and
$\, W_{\alpha} \, $ are simply connected (\cite{UltraCompl}). 
This, in particular, implies the relations
$$\max_{W_{\alpha}} \, \widetilde{\epsilon}_{1} ({\bf n}) 
\,\, = \,\, 
\max_{\partial W_{\alpha}} \, \widetilde{\epsilon}_{1} ({\bf n}) 
\,\,\, , \quad \quad
\min_{W_{\alpha}} \, \widetilde{\epsilon}_{2} ({\bf n}) 
\,\, = \,\, 
\min_{\partial W_{\alpha}} \, \widetilde{\epsilon}_{2} ({\bf n}) $$
for all Zones $\, W_{\alpha} \, $ (and their boundaries 
$\, \partial W_{\alpha}$). The relations (\ref{eB12}) can then be 
replaced by the relations
\begin{equation}
\label{eBprime}
\epsilon^{\cal B}_{1} \,\,\, = \,\,\, \min^{\prime} \, 
\widetilde{\epsilon}_{2} ({\bf n}) \quad , \quad \quad
\epsilon^{\cal B}_{2} \,\,\, = \,\,\, \max^{\prime} \, 
\widetilde{\epsilon}_{1} ({\bf n}) \,\,\, , 
\end{equation}
where the minimum and maximum are taken only over the set of 
boundaries of $\, W_{\alpha} \, $ (and not over the entire 
unit sphere $\, \mathbb{S}^{2}$).

 On the boundaries of $\, W_{\alpha} \, $ we have
$$\widetilde{\epsilon}_{1} ({\bf n}) \,\,\, = \,\,\, 
\widetilde{\epsilon}_{2} ({\bf n}) \,\,\, \equiv \,\,\,
\widetilde{\epsilon}_{0} ({\bf n}) \,\,\, , $$
and thus the position of the interval 
$\, \left[ \epsilon^{\cal B}_{1} , \, \epsilon^{\cal B}_{2} \right] \, $ 
is determined by the minimum and maximum of the function 
$\, \widetilde{\epsilon}_{0} ({\bf n}) \, $ over all boundaries 
of $\, W_{\alpha} \, $:
$$\epsilon^{\cal B}_{1} \,\,\, = \,\,\, 
\min_{\cup \partial W_{\alpha}} \, 
\widetilde{\epsilon}_{0} ({\bf n}) \quad , \quad \quad
\epsilon^{\cal B}_{2} \,\,\, = \,\,\, 
\max_{\cup \partial W_{\alpha}} \, 
\widetilde{\epsilon}_{0} ({\bf n}) $$

 For ``physically realistic'' relations $\, \epsilon ({\bf p}) \, $, 
the location of $\, \epsilon_{F} \, $ in the interval 
$\, \left( \epsilon^{\cal B}_{1} , \, \epsilon^{\cal B}_{2} \right) \, $ 
entails a significant complication of conductivity diagrams compared to 
the case 
$\, \epsilon_{F} \notin \left( \epsilon^{\cal B}_{1} , 
\, \epsilon^{\cal B}_{2} \right) \, $. 
In particular, the regions of electron and hole Hall conductivity 
(and the presence of only closed trajectories on the Fermi surface) 
are divided on $\, \mathbb{S}^{2} \, $ by ``quasi-one-dimensional'' 
chains consisting of Zones $\, \Omega_{\alpha} \, $ (Fig. \ref{Fig5}, c, d). 
The number of Zones $\, \Omega_{\alpha} \subset \mathbb{S}^{2} \, $ 
for generic values 
$\, \epsilon_{F} \in \left[ \epsilon^{\cal B}_{1} , \, 
\epsilon^{\cal B}_{2} \right] \, $ is infinite, which entails 
the emergence of stable open trajectories of arbitrarily complex 
shape (Fig. \ref{Fig7}, a) for certain directions of $\, {\bf B} \, $. 
The accumulation points of Zones $\, \Omega_{\alpha} \, $ correspond 
here, as a rule, to the emergence of chaotic trajectories 
(of the Tsarev or Dynnikov type) on the Fermi surface.

\vspace{1mm}

 Finding the functions $\, \widetilde{\epsilon}_{0} ({\bf n}) \, $ 
is a complex problem requiring rather nontrivial calculations. Since 
we use here only some general approximation for the relations 
$\, \epsilon ({\bf p}) \, $, our goal here will be to determine 
the interval 
$\, \left[ \epsilon^{\cal B}_{1} , \, \epsilon^{\cal B}_{2} \right] \, $ 
with some (good) accuracy. For this purpose, we use here a technique 
proposed in the paper \cite{TightBind}, which gives good 
(and in some cases accurate) estimates for the quantities 
$\, \epsilon^{\cal B}_{1} \, $ and $\, \epsilon^{\cal B}_{2} \, $.

 As we have already said, the interval 
$\, \left[ \epsilon^{\cal B}_{1} , \, \epsilon^{\cal B}_{2} \right] \, $ 
is determined in our case by the values of 
$\, \widetilde{\epsilon}_{0} ({\bf n}) \, $ 
on the boundaries of the Zones $\, W_{\alpha} \, $, i.e. on the set
$$\bigcup_{\alpha} \partial W_{\alpha} $$

 For approximate evaluation of the interval
$\left[ \epsilon^{\cal B}_{1} , \, \epsilon^{\cal B}_{2} \right]$, 
we can consider the values $\, \widetilde{\epsilon}_{0} ({\bf n}) \, $ 
on the boundary of one of the largest Zones $\, W_{\alpha} \, $, 
where the interval of values of $\, \widetilde{\epsilon}_{0} ({\bf n}) \, $ 
is sufficiently wide. For symmetric relations $\, \epsilon ({\bf p}) \, $ 
such Zones are, as a rule, symmetric Zones $\, W_{\alpha} \, $ 
(the existence of which was proved in \cite{dynn3}). To calculate 
the minimum and maximum of $\, \widetilde{\epsilon}_{0} ({\bf n}) \, $ 
on the boundary of such Zones, it is usually sufficient to calculate 
$\, \widetilde{\epsilon}_{0} ({\bf n}) \, $ at the most ``symmetric'' 
points $\, \partial W_{\alpha} \, $ (Fig. \ref{Fig8}).

\vspace{1mm}

\begin{figure*}[t]
\begin{center}
\includegraphics[width=\linewidth]{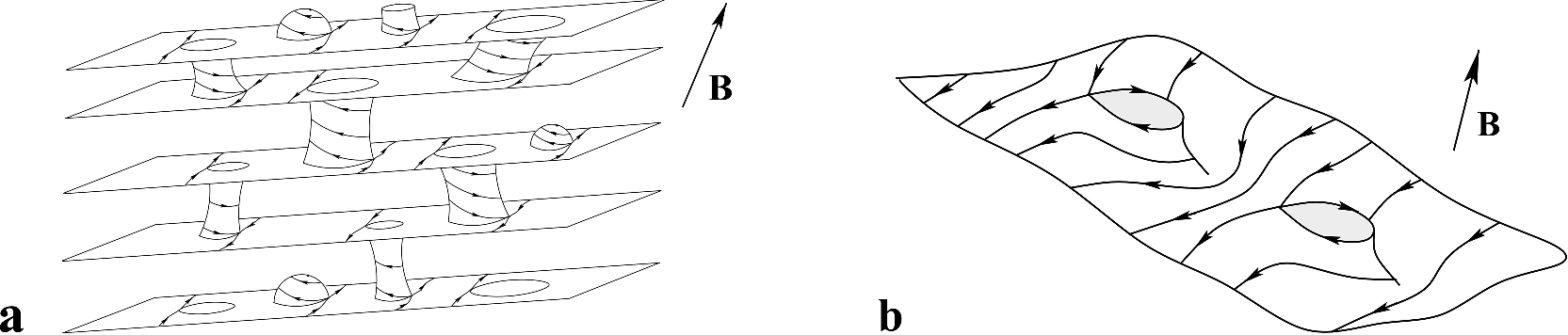}
\end{center}
\caption{(a) The Fermi surface cut by the cylinders of closed 
trajectories of (\ref{MFSyst}) (schematically). 
(b) A carrier of stable open trajectories of system (\ref{MFSyst}) 
in the full $\, {\bf p}$ - space.}
\label{Fig9}
\end{figure*}

 To determine the boundaries of the Zones $\, W_{\alpha} \, $ 
(as well as the values $\, \widetilde{\epsilon}_{0} ({\bf n}) $), 
it is necessary to consider the topological structure of system 
(\ref{MFSyst}) on the surfaces
$$\epsilon ({\bf p}) \,\,\, = \,\,\, {\rm const}  \,\,\, , $$
corresponding to the emergence of stable open trajectories 
(\cite{zorich1,dynn1,dynn3}). Namely, for any
$${\bf n} \, \in \, W_{\alpha} \,\,\, , \quad
\epsilon_{F} \, \in \, \left[ \widetilde{\epsilon}_{1} ({\bf n}) ,
\widetilde{\epsilon}_{2} ({\bf n}) \right] $$
the surface $\, \epsilon ({\bf p}) = \epsilon_{F} \, $ 
represents a set of ``carriers of open trajectories'' separated by 
cylinders of closed trajectories of (\ref{MFSyst}) 
(Fig. \ref{Fig9}, a).

\begin{figure*}[t]
\begin{center}
\includegraphics[width=\linewidth]{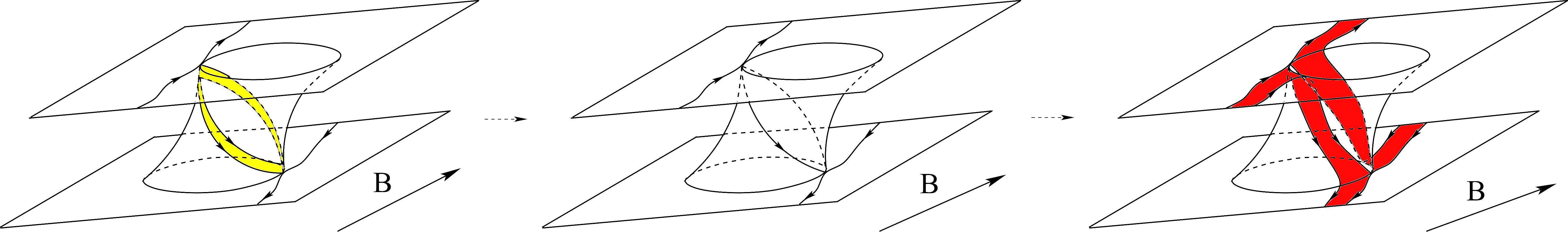}
\end{center}
\caption{The vanishing of the height of a cylinder of closed 
trajectories at the boundary of a Zone $\, \Omega_{\alpha} \, $ 
and the destruction of a ``stable'' topological structure of system 
(\ref{MFSyst}) on the Fermi surface (the occurrence of ``jumps'' 
between the carriers of open trajectories).}
\label{Fig10}
\end{figure*}

 The carriers of stable open trajectories are periodically 
deformed integer planes (with holes) parallel to each other 
in $\, \mathbb{R}^{3} \, $ (Fig. \ref{Fig9}, b) 
(\cite{zorich1,dynn1,dynn3}). A change in the topological 
structure of (\ref{MFSyst}) on the surface $\, S_{F} \, $ for 
directions $\, {\bf n} \in \partial \Omega_{\alpha} \, $ corresponds 
to the vanishing of the height of one of the cylinders of closed 
trajectories separating the carriers of open trajectories 
(Fig. \ref{Fig10}).

\vspace{1mm}

\vspace{1mm}

 The boundaries of Zones $\, W_{\alpha} \, $ correspond to 
the disappearance of at least two cylinders of closed trajectories 
of (\ref{MFSyst}), which have different (electron and hole) types 
(\cite{dynn3}). The condition of the simultaneous disappearance of 
two cylinders of closed trajectories for
$${\bf n} \, \in \, \partial W_{\alpha} \,\,\, , \quad \quad
\epsilon_{F} \, = \, \widetilde{\epsilon}_{0} ({\bf n}) $$
determines both the boundary of a Zone $\, W_{\alpha} \, $ and 
the values $\, \widetilde{\epsilon}_{0} ({\bf n}) \, $.

\vspace{1mm}

 In general, studying the structure of system (\ref{MFSyst}) 
on a Fermi surface is quite non-trivial and requires extensive 
computations (see, e.g., \cite{DynnBuDA,DeLeo2017,DynMalNovUMN}). 
This study is significantly simplified for symmetric Zones 
$\, W_{\alpha} \, $, where such a structure becomes more obvious. 
The use of ``symmetric'' points on the boundaries of such Zones 
makes it most convenient to find both their 
positions on $\, \mathbb{S}^{2} \, $ and the corresponding 
values $\, \widetilde{\epsilon}_{0} (P, Q) \, $.

 Below we illustrate our scheme using the example of the simple 
cubic lattice, which represents the simplest of the cases considered 
here.

\vspace{1mm}

 The case of the simple cubic lattice is the simplest from the geometric 
point of view. The reciprocal lattice here is also simple cubic; we will 
assume its edge length to be equal to 2. As we have already said, we will 
consider the dispersion relations $\, \epsilon ({\bf p}) \, $, obtained 
in the nearly free electron approximation. 

 Our goal is to determine the position of the interval 
$\, \left[ \epsilon^{\cal B}_{1} , \, \epsilon^{\cal B}_{2} \right] \, $ 
within the interval 
$\, \left[ \epsilon_{\min} , \, \epsilon_{\max} \right] \, $ 
and to estimate its width in comparison with the conduction band width. 
Based on this, we can normalize the values of $\, \epsilon \, $ 
and $\, p \, $ such that
$$\epsilon \,\,\, = \,\,\, p^{2} $$
near the value $\, {\bf p} = (0, 0, 0) \, $.

\begin{figure*}[t]
\begin{center}
\includegraphics[width=\linewidth]{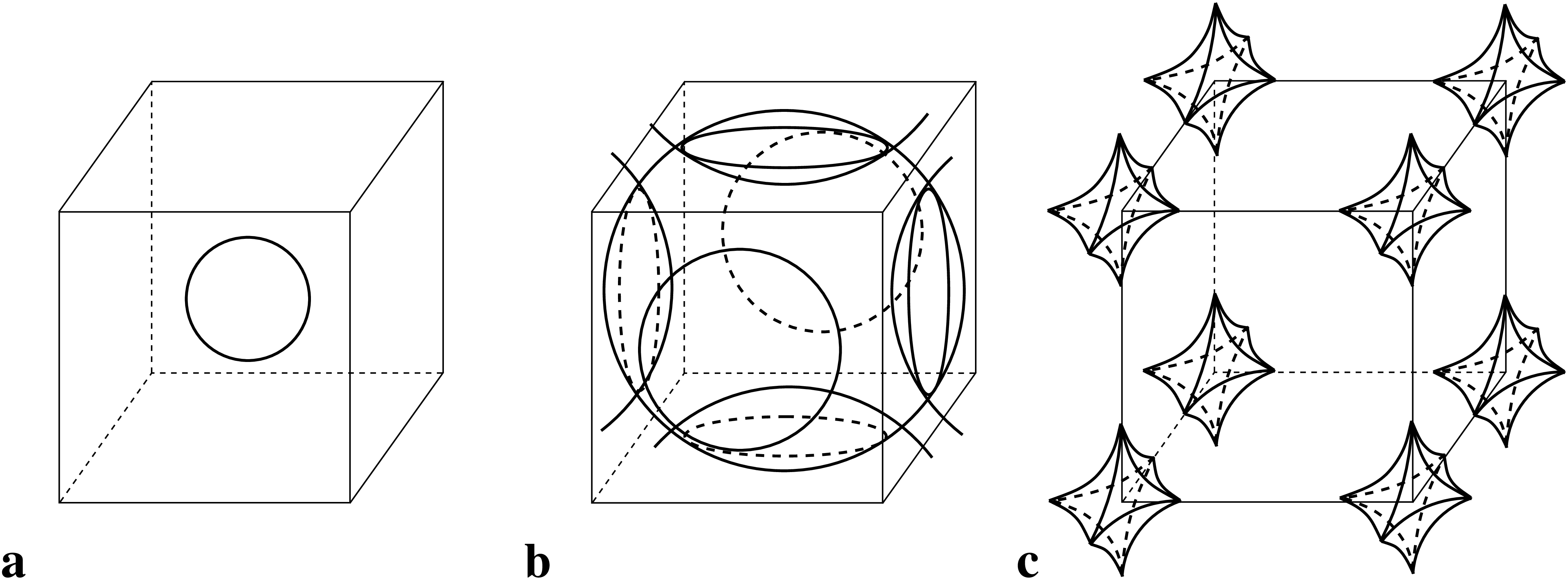}
\end{center}
\caption{(a) Emergence of the components of the Fermi surface of 
the first conduction band at $\, \epsilon_{F} > 0 \, $.
(b) Growth and surgery of the Fermi surface of the first 
conduction band at $\, \epsilon_{F} > 0 \, $.
(c) Disappearance of the components of the Fermi surface of 
the first conduction band at $\, \epsilon \rightarrow 3 \, $ 
(simple cubic lattice).}
\label{Fig11}
\end{figure*}

 The evolution of the Fermi surface for $\, \epsilon_{F} > 0 \, $ 
begins with the emergence of growing spheres with centers at even 
integer points (Fig. \ref{Fig11}, a) in $\, \mathbb{R}^{3} \, $ 
with their subsequent surgery upon the formation of intersections 
(Fig. \ref{Fig11}, b).

 The Fermi surface first becomes extended in the entire 
$\, {\bf p}$ - space at $\, \epsilon_{F} = 1 \, $, this is also 
the value after which stable open trajectories of system 
(\ref{MFSyst}) ($\epsilon^{\cal A}_{1} = 1$) first appear on it.

 For values 
$\, \epsilon_{F} \in (1, 2) \, $ ($p_{F} \in (1, \sqrt{2})$), 
the Fermi surface contains a connected component extending in all 
directions in $\, {\bf p}$ - space, as well as a periodic set of 
compact components (Fig. \ref{Fig11}, b). It is easy to see that 
among the compact components there are only 3 nonequivalent ones, 
and the total number of connected components of the Fermi surface 
in $\, \mathbb{T}^{3} = \mathbb{R}^{3} / L^{*} \, $ is equal to 4.

 In fact, the emergence of new components of the Fermi surface 
corresponds to the creation of the second conduction band with 
$\, \epsilon^{(2)}_{\min} = 1 \, $. In the interval 
$\, \epsilon \in (1, 2) \, $ we must therefore consider the 
unbounded component to be the Fermi surface of the first conduction 
band (with $\, \epsilon^{(1)}_{\min} = 0 $), and the 3 compact 
components to be the Fermi surface of the second band 
(with $\, \epsilon^{(2)}_{\min} = 1 $).

 At $\, \epsilon_{F} = 2 \, $, the Fermi surface of the first band 
decomposes into compact components of rank 0 (equivalent to each other), 
which disappear (contract into points) at $\, \epsilon_{F} = 3 \, $ 
(Fig. \ref{Fig11}, c). Thus, for the first conduction band we have
$$\epsilon^{(1)}_{\min} = 0 \quad , \quad \quad \epsilon^{(1)}_{\max} = 3 $$

\vspace{1mm}

 The Fermi surface of the second band acquires rank 3 in the interval 
$\, \epsilon \in (2, 3) \, $ ($p \in (\sqrt {2}, \sqrt {3})$), extending 
in all directions in $\, {\bf p}$ - space. For $\, \epsilon = 3 \, $ 
the Fermi surface of the second band decomposes into components 
of rank zero, which vanish at the value $\, \epsilon = 4 \, $ ($p = 2$). 
We thus have for the second conduction band
$$\epsilon^{(2)}_{\min} = 1 \quad , \quad \quad \epsilon^{(2)}_{\max} = 4 $$

 At the same time, for $\, \epsilon = 2 \, $ ($p = \sqrt{2}$), 
new components of rank 0 are split off from the Fermi surface of 
the second conduction band, forming the Fermi surfaces of the 
following conduction bands, etc. (according to \cite{Harrison}, 
each point in $\, {\bf p}$ - space that falls inside $\, n \, $ 
different spheres for a given $\, \epsilon_{F} \, $ represents 
filled states for the $\, n \, $ first conduction bands). 
Continuing the procedure, we thus obtain an infinite set 
of (overlapping) energy bands (conduction bands).

\vspace{1mm}

 At all points of $\, {\bf p}$ - space we have the relations
$$\epsilon^{(s+1)} ({\bf p}) \,\,\, \geq \,\,\, 
\epsilon^{(s)} ({\bf p}) \,\,\, , $$
at the same time
$$\epsilon^{(s+1)} ({\bf p}) \,\,\, = \,\,\, \epsilon^{(s)} ({\bf p}) $$
at the intersections of the Fermi surfaces.
 
 As is well known, in real nearly free electron approximation,
the last equality is removed due to the (small) splitting of energy 
levels at the intersection of spectra. The Fermi surfaces are then 
smoothed out at the intersections, leading to the relations
$$\epsilon^{(s+1)} ({\bf p}) \,\,\, > \,\,\, 
\epsilon^{(s)} ({\bf p}) $$

 Our results here, however, are approximate (and, in particular, 
do not take into account deviations from the nearly free-electron 
approximation in real metals). Therefore, we will also consider 
here the shape of the Fermi surfaces as described above.

\vspace{1mm}

 Here we will estimate the position of the interval 
$\, \left[ \epsilon^{{\cal B}}_{1} , \, 
\epsilon^{{\cal B}}_{2} \right] \, $ 
for the first conduction band described above. As we have already 
said, we use here the values 
$\, \widetilde{\epsilon}_{0} ({\bf n}) \, $ at those points of
$\, \partial W_{\alpha} \, $ where we can expect them to differ from 
each other the most. As such values, we can use the values 
$\, \widetilde{\epsilon}_{1} ({\bf n}) \, = \, 
\widetilde{\epsilon}_{2} ({\bf n}) \, = \, 
\widetilde{\epsilon}_{0} ({\bf n}) \, $ 
on the boundaries of the largest (symmetric) Stability Zones 
$\, W_{\alpha} \subset \mathbb{S}^{2} \, $. Here (and further) 
we will use the symmetric Zone $\, W_{1} \, $, containing the direction 
$\, {\bf n} = (0, 0, 1) \, $ (Fig. \ref{Fig8}).

\begin{figure}[t]
\begin{center}
\includegraphics[width=0.8\linewidth]{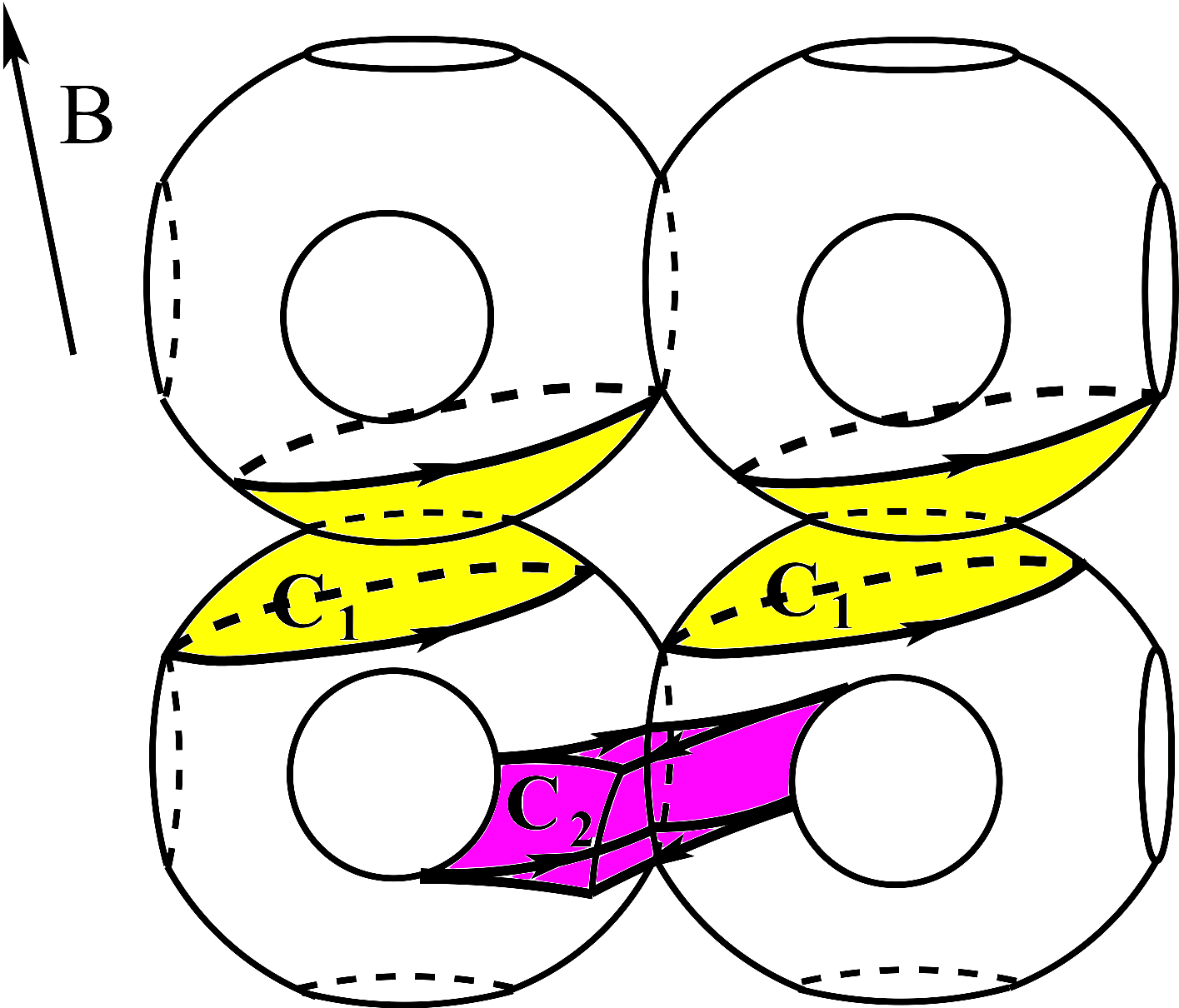}
\end{center}
\caption{The Fermi surface of the first conduction band for 
$\, \epsilon_{F} \in (1, 2) \, $ and the cylinders of closed 
trajectories of the electron ($C_{1}$) and hole ($C_{2}$) types 
for $\, {\bf n} \in W_{1} \, $ (simple cubic lattice).}
\label{Fig12}
\end{figure}

 The boundary points of the Zone $\, W_{1} \, $ are determined by 
the disappearance of two cylinders of closed trajectories of 
(\ref{MFSyst}), dividing the Fermi surface into carriers of open 
trajectories for $\, {\bf n} \in W_{1} \, $. For the first conduction 
band, we obviously have
$$\epsilon^{{\cal A}}_{1} \,\,\, = \,\,\, 1 \quad  ,
\quad \quad \epsilon^{{\cal A}}_{2} \,\,\, = \,\,\, 2 
\quad , $$
so that it suffices to consider the Fermi surfaces only for 
$\, \epsilon_{F} \in (1, 2) \, $. For $\, {\bf n} \in W_{1} \, $, 
the corresponding cylinders of closed trajectories are shown 
in Fig. \ref{Fig12}. To estimate the values
$\, \epsilon^{{\cal B}}_{1} \, $ and $\, \epsilon^{{\cal B}}_{2} \, $ 
we use here the values of $\, \widetilde{\epsilon}_{0} ({\bf n}) \, $ 
at ``symmetric'' points $\, P \, $ and $\, Q \, $ on the boundary of 
$\, W_{1} \, $ shown in Fig. \ref{Fig8}.

\vspace{1mm}

\begin{figure*}[t]
\begin{center}
\includegraphics[width=0.9\linewidth]{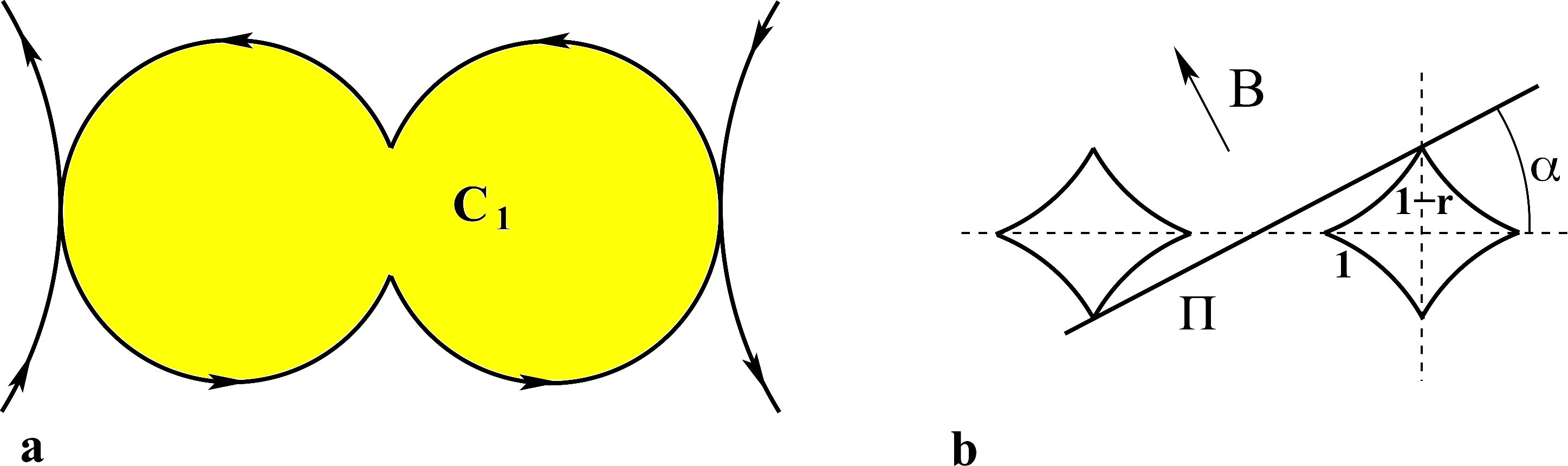}
\end{center}
\caption{(a) Cylinder $\, C_{1} \, $ of zero height in the plane 
$\, \Pi \perp {\bf B} \, $ passing through the point 
$\, (0, 0, 1) \, $ (${\bf n} = P$); 
(b) Projection of the plane $\, \Pi \, $ onto the plane $\, yz \, $ 
(simple cubic lattice).}
\label{Fig13}
\end{figure*}

 Let the radius of the circle arising at the intersection of 
the spheres for $\, 1 < \epsilon_{F} < 2 \, $ be $\, r \, $. 
The disappearance of the cylinder $\, C_{1} \, $ at the point 
$\, P \, $ corresponds to the picture in the plane 
$\, \Pi \perp {\bf B} \, $ (passing through the point 
$\, (0, 0, 1) \in \mathbb{R}^{3}$) shown in Fig. \ref{Fig13}, a. 
The projection of the plane $\, \Pi \, $ onto the plane $\, yz \, $ 
is a straight line with an inclination angle $\, \alpha \, $ 
such that
$$\tan \alpha \,\,\, = \,\,\, 1 \, - \, r $$
(Fig. \ref{Fig13}, b).

\begin{figure*}[t]
\begin{center}
\includegraphics[width=0.9\linewidth]{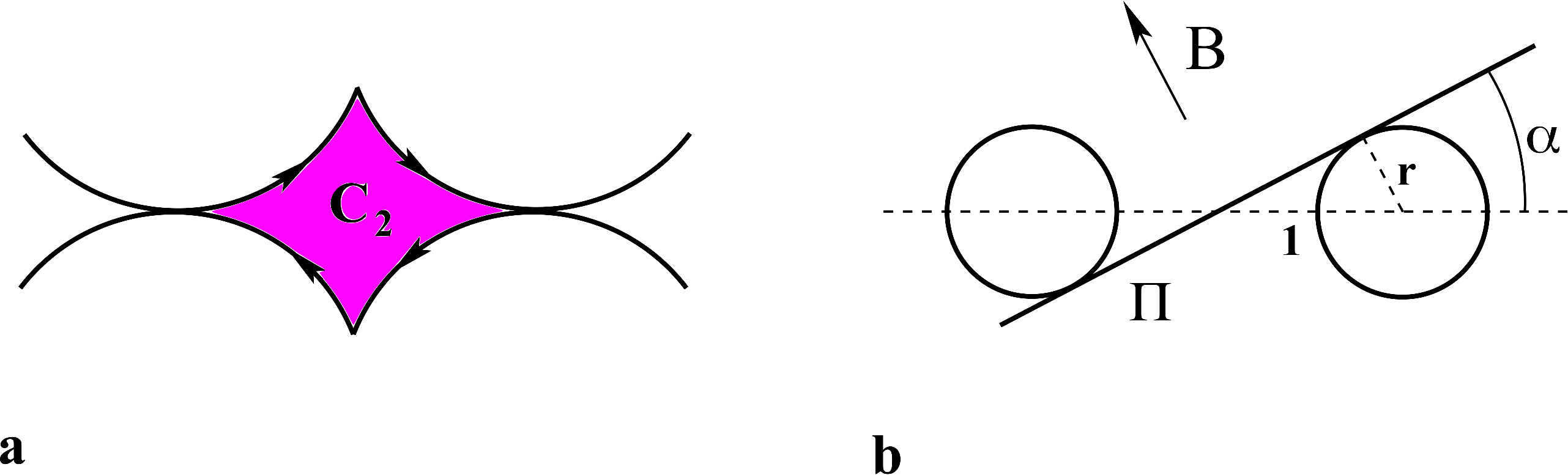}
\end{center}
\caption{(a) Cylinder $\, C_{2} \, $ of zero height in the plane 
$\, \Pi \perp {\bf B} \, $ passing through the point 
$\, (1, 1, 0) \, $ (${\bf n} = P$); 
(b) Projection of the plane $\, \Pi \, $ onto the plane 
$\, yz \, $ (simple cubic lattice).}
\label{Fig14}
\end{figure*}

 The disappearance of the cylinder $\, C_{2} \, $ at the point 
$\, P \, $ corresponds to the picture in the plane 
$\, \Pi \perp {\bf B} \, $ (passing through the point 
$\, (1, 1, 0) \in \mathbb{R}^{3}$), shown in Fig. \ref{Fig14}, a. 
The projection of the plane $\, \Pi \, $ onto the plane $\, yz \, $ 
is a straight line with an inclination angle $\, \alpha \, $ such that
$$\sin \alpha \,\,\, = \,\,\, r $$
(Fig. \ref{Fig14}, b).

 From the condition of the simultaneous disappearance of the 
cylinders $\, C_{1} \, $ and $\, C_{2} \, $
$$1 - r \,\,\, = \,\,\, {r \over \sqrt{1 - r^{2}}} $$
we then obtain $\, r \simeq 0.469 \, $ and, accordingly,
$$\widetilde{\epsilon}_{0} (P) \,\,\, = \,\,\, 1 + r^{2} 
\,\,\, \simeq \,\,\, 1.22 $$

\vspace{1mm}

\begin{figure}[t]
\begin{center}
\includegraphics[width=\linewidth]{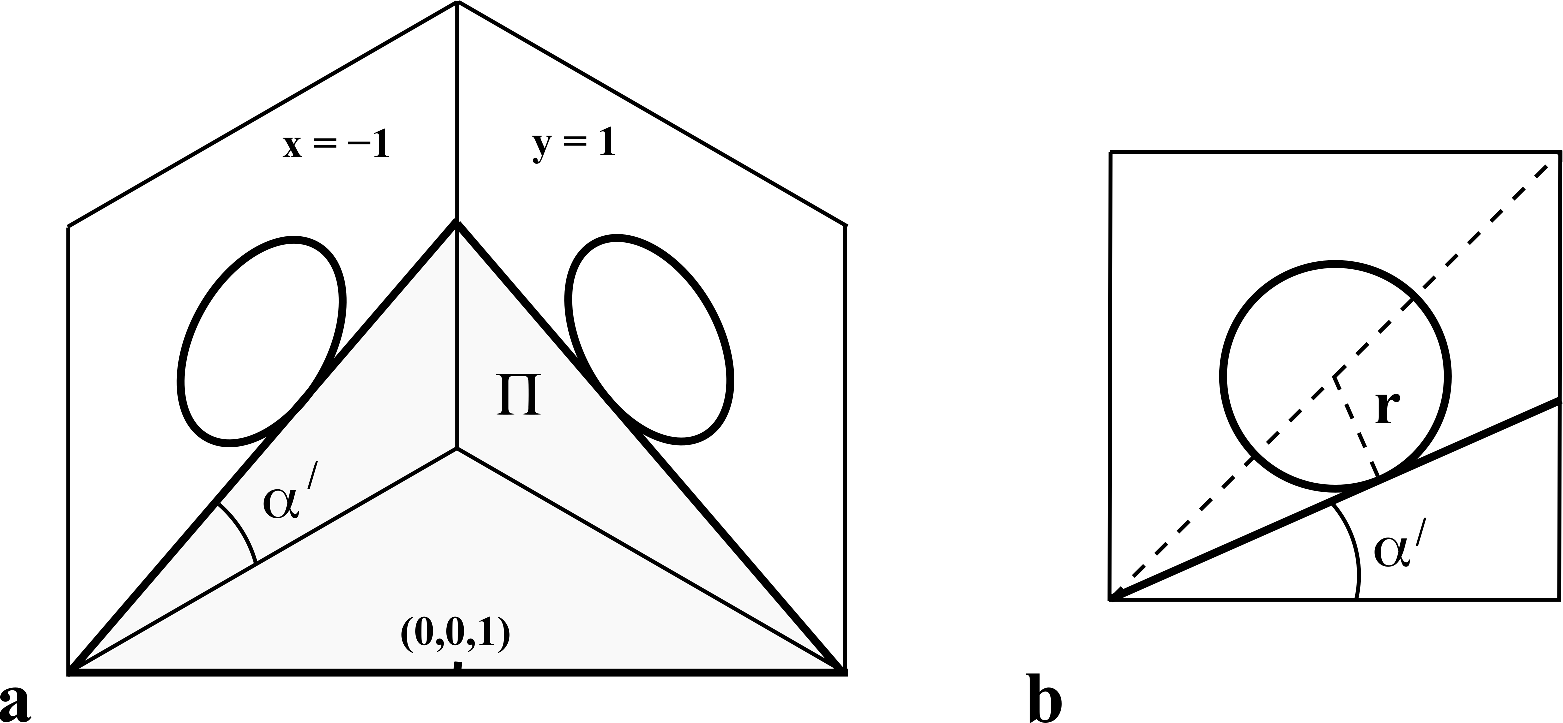}
\end{center}
\caption{(a) The position of the plane $\, \Pi \perp {\bf B} \, $ 
passing through the point $\, (0, 0, 1) \, $ at $\, {\bf n} = Q\, $ 
(cube interior shown);
(b) The intersection of the plane $\, x = - 1 \, $ by the plane 
$\, \Pi \, $ and angle $\, \alpha^{\prime} \, $ 
($\tan \alpha = \sqrt{2} \tan \alpha^{\prime}$, simple cubic lattice).}
\label{Fig15}
\end{figure}

 The disappearance of the cylinder $\, C_{1} \, $ at 
$\, {\bf n} = Q \, $ corresponds to the position of the plane 
$\, \Pi \perp {\bf B} \, $ (passing through the point 
$\, (0, 0, 1) \in \mathbb{R}^{3}$), shown in Fig. \ref{Fig15}, a. 
As is easy to see, the angle between the directions $\, {\bf n} \, $ 
and $\, z \, $ is determined by the relation
$$\tan \alpha \,\,\, = \,\,\, \sqrt{2} \, \tan \alpha^{\prime} \,\,\, , $$
where $\, \alpha^{\prime} \, $ is the angle shown in Fig. \ref{Fig15}, b. 
As is easy to show,
$$\tan \alpha^{\prime} \,\,\, = \,\,\, 
{\sqrt{2 - r^{2}} - r \over \sqrt{2 - r^{2}} + r} $$

\vspace{1mm}

\begin{figure}[t]
\begin{center}
\includegraphics[width=\linewidth]{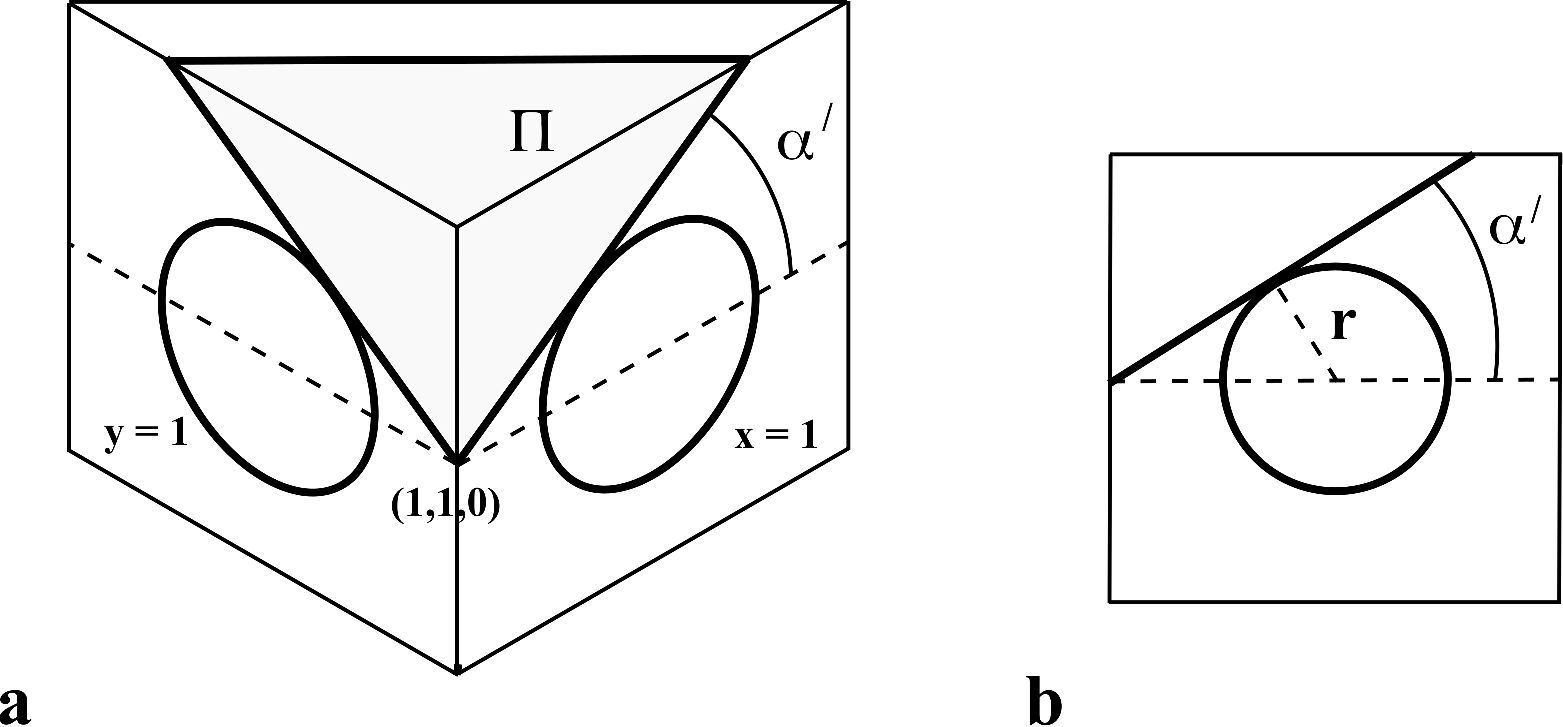}
\end{center}
\caption{(a) The position of the plane $\, \Pi \perp {\bf B} \, $ 
passing through the point $\, (1, 1, 0) \, $ at $\, {\bf n} = Q\, $ 
(the exterior of the cube is shown);
(b) The intersection of the plane $\, x = 1 \, $ by the plane 
$\, \Pi \, $ and the angle $\, \alpha^{\prime} \, $ 
($\tan \alpha = \sqrt{2} \tan \alpha^{\prime}$, simple cubic lattice).}
\label{Fig16}
\end{figure}

 The disappearance of the cylinder $\, C_{2} \, $ at 
$\, {\bf n} = Q \, $ corresponds to the position of the plane 
$\, \Pi \perp {\bf B} \, $ (passing through the point 
$\, (1, 1, 0) \in \mathbb{R}^{3}$), shown in Fig. \ref{Fig16}, a. 
The angle between the directions $\, {\bf n} \, $ and $\, z \, $ 
is determined by the relation
$$\tan \alpha \,\,\, = \,\,\, \sqrt{2} \, \tan \alpha^{\prime} 
\,\,\, = \,\,\, \sqrt{2} \, {r \over \sqrt{1 - r^{2}}} $$

 The condition
$${\sqrt{2 - r^{2}} - r \over \sqrt{2 - r^{2}} + r} 
\,\,\, = \,\,\, {r \over \sqrt{1 - r^{2}}} $$
gives the value $\, r = 1 / \sqrt{5} \,\, $ and, accordingly,
$$\widetilde{\epsilon}_{0} (Q) \,\,\, = \,\,\, 1 + r^{2} 
\,\,\, = \,\,\, 1.2 $$

 Thus, using our estimate, we obtain
$$\left[ \epsilon^{\cal B}_{1} , \, \epsilon^{\cal B}_{2} \right] 
\,\,\, \simeq \,\,\, \left[ 1.2 , \, 1.22 \right] $$
for the first conduction band.

 Abstracting from the normalization we use, we can also say that 
the interval of emergence of ultra-complex angular diagrams occurs 
near the value $\, 0.4 \, \epsilon_{\rm \max} \, $ from the bottom 
of the conduction band, and its length is $\, \simeq \, 0.7 \% \, $ 
of the width of the conduction band.

\vspace{1mm}

 Note also that the tight-binding approximation gives the zero width
of the interval 
$\, \left[ \epsilon^{\cal B}_{1} , \epsilon^{\cal B}_{2} \right] \, $ 
in the leading order for the simple cubic lattice 
(see e.g. \cite{TightBind}), and a nonzero width of 
$\, \left[ \epsilon^{\cal B}_{1} , \epsilon^{\cal B}_{2} \right] \, $ 
appears there only in higher corrections. The nearly free electron 
approximation, as can be seen, differs from the tight-binding model 
at this point.

\section{The face-centered cubic lattice}
\setcounter{equation}{0}

 Let's consider the case of the face-centered cubic lattice. 
The reciprocal lattice in this case is body-centered cubic. 
As in the previous case, we will assume its edge length to be 
equal to 2.

 The first Fermi surfaces are born near integer points 
$\, {\bf p} = (n, m, l) \, $, all of whose coordinates are either 
even or odd. Near these points, as in the previous section,
we will assume
$$\epsilon ({\bf p}) \,\,\, = \,\,\, \left( \Delta {\bf p} \right)^{2} $$

 As in the previous case, we will carry out here an estimate of the 
position of the interval 
$\, \left[ \epsilon^{\cal B}_{1} , \, \epsilon^{\cal B}_{2} \right] \, $
for the first conduction band.

\begin{figure*}[t]
\begin{center}
\includegraphics[width=0.9\linewidth]{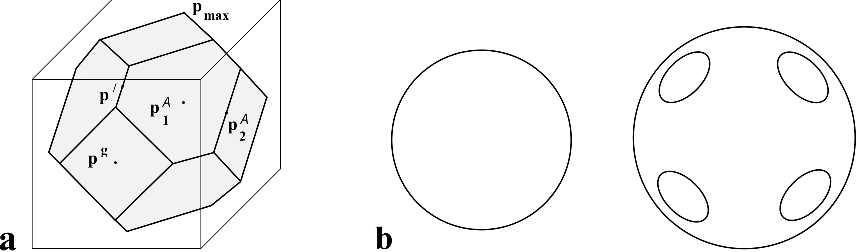}
\end{center}
\caption{(a) Brillouin zone for the face-centered cubic lattice 
(reciprocal lattice - body-centered cubic).
(b) Surgery of the Fermi surface upon passing through 
$\epsilon_{F} = 3/4$.}
\label{Fig17}
\end{figure*}

 As is well known, the Brillouin zone in this case is a truncated 
octahedron and has the shape shown in Fig. \ref{Fig17}, a. The bottom 
of the first conduction band in our normalization obviously corresponds 
to the value $\, \epsilon_{\rm min} = 0 \, $. The upper edge of the 
first conduction band corresponds to the values
$$\epsilon_{\rm max} \,\,\, = \,\,\, {13 \over 9} \,\,\, ,
\quad \quad p_{\rm max} \,\,\, = \,\,\,
\sqrt{1 + \left({2 \over 3}\right)^{2}} $$
(Fig. \ref{Fig17}, a).

 Up to the value
$$\epsilon_{F} \,\,\, = \,\,\, \epsilon^{\cal A}_{1} 
\,\,\, = \,\,\, {3 \over 4} $$
the Fermi surface is a sphere. At 
$\, \epsilon_{F} = \epsilon^{\cal A}_{1} = 3/4 \, $ 
the spheres first touch each other 
($p^{\cal A}_{1} = \sqrt{3}/2 \, $, Fig. \ref{Fig17}, b), 
the Fermi surface acquires genus 4, and open trajectories 
first appear on it.

 After the value
$$\epsilon_{F} \,\,\, = \,\,\, \epsilon^{g} \,\,\, = \,\,\, 1 $$
($p^{g} = 1 \, $) the genus of the Fermi surface increases to 7.

 After the value
$$\epsilon_{F} \,\,\, = \,\,\, \epsilon^{\cal A}_{2} 
\,\,\, = \,\,\, {11 \over 9} \,\,\, ,  \quad \quad 
p^{\cal A}_{2}  \,\,\, = \,\,\,
\sqrt{1 + \left({\sqrt{2} \over 3}\right)^{2}} $$
the Fermi surface breaks down into 6 compact components that 
look like ``dumbbells''.

 After the value
$$\epsilon_{F} \,\,\, = \,\,\, \epsilon^{\prime} 
\,\,\, = \,\,\, {25 \over 18} \,\,\, ,  \quad \quad 
p^{\prime}  \,\,\, = \,\,\, {5 \over 6} \, \sqrt{2} $$
the ``dumbbells'' disintegrate into ``balls'' of the hole type, 
which then disappear at
$\, \epsilon_{F} = \epsilon_{\rm max} \, $. 

\vspace{1mm}

\begin{figure*}[t]
\begin{center}
\includegraphics[width=\linewidth]{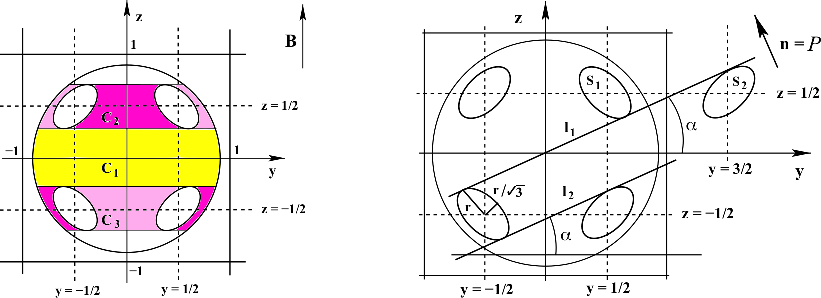}
\end{center}
\caption{Projection of the Fermi surface onto the $\, y z \, $ 
plane for $\, {\bf B} \parallel z \, $ and disappearance of the 
cylinders $\, C_{1} \, $, $\, C_{2} \, $ and $\, C_{3} \, $ for 
$\, {\bf n} = P \, $ (face-centered lattice).}
\label{Fig18}
\end{figure*}

 Let us now consider the projection of our Fermi surface onto the 
$\, y z \, $ plane. The circles arising at the intersection of the 
spheres are now projected onto ellipses with semi-axes 
$\, r \, $ and $\, r / \sqrt{3} \, $ (Fig. \ref{Fig18}). 
It is also easy to construct projections of the cylinders 
$\, C_{1} \, $, $\, C_{2} \, $ and $\, C_{3} \, $ onto the 
$\, y z \, $ plane for $\, {\bf B} \parallel z \, $ (Fig. \ref{Fig18}). 
As in the previous section, the cylinders $\, C_{1} \, $, $\, C_{2} \, $, 
and $\, C_{3} \, $ cut the Fermi surface into carriers of open trajectories 
$\, \mathbb{T}^{2}_{1} \, $, $\, \mathbb{T}^{2}_{2} \, $ for 
$\, {\bf n} \in W_{1} \, $. Note here that the cylinder $\, C_{1} \, $ 
has a circular shape, while the cylinders $\, C_{2} \, $ and $\, C_{3} \, $ 
have ``corners'', like the cylinder $\, C_{2} \, $ in the previous section.

 As can be seen, the disappearance of the cylinders 
$\, C_{1} \, $, $\, C_{2} \, $ and $\, C_{3} \, $ for the direction 
$\, {\bf n} = P \, $ (Fig. \ref{Fig8}) corresponds to the situation 
when the straight line passing through the point $\, (0, 0) \, $ 
(the projection of the plane orthogonal to $\, {\bf B}$) is tangent 
to both ellipses $\, S_{1} \, $ and $\, S_{2} \, $ shown 
in Fig. \ref{Fig18}. This condition, as is easy to see, uniquely 
determines the quantity $\, r \, $, and also the value  
$\, \widetilde{\epsilon}_{0} (P) \, = \, 3/4 + r^{2} \, $.

 It is easy to verify that the equations of the ellipses 
$\, S_{1} \, $ and $\, S_{2} \, $ shown in Fig. \ref{Fig18} 
have the form
$$\left( y - z \right)^{2} \,\, + \,\, 
3 \left( y + z - 1 \right)^{2} \,\,\, = \,\,\, 2 \, r^{2} $$
and
$$3 \left(y - z - 1 \right)^{2} \,\, + \,\, 
\left( y + z - 2 \right)^{2} \,\,\, = \,\,\, 2 \, r^{2} $$
respectively.

 The condition of tangency of the ellipse $\, S_{1} \, $ is
$$2 \left( 1 - \tan \alpha \right)^{2}  r^{2} \,\, + \,\, 
6 \left( 1 + \tan \alpha \right)^{2} r^{2} 
\,\,\, = \,\,\, 3 \left( 1 - \tan \alpha \right)^{2}  , $$
and of the ellipse $\, S_{2} \, $ is
$$6 \left( 1 - \tan \alpha \right)^{2}  r^{2} \,\, + \,\, 
2 \left( 1 + \tan \alpha \right)^{2} r^{2} 
\,\,\, = \,\,\, 3 \left( 3 \tan \alpha - 1 \right)^{2}  , $$
which gives
$$r^{2} (P) \,\,\, \simeq \,\,\, 0.066885 $$
and
$$\widetilde{\epsilon}_{0} (P) \,\,\, \simeq \,\,\, 0.816885 $$

 For the angle of inclination of $\, {\bf B} \, $ to the axis 
$\, z \, $, corresponding to the point $\, P \, $ on the 
angular diagram, we have
$$\tan \alpha_{P} \,\,\, \simeq \,\,\, 0.45541 $$

\vspace{2mm}

\begin{figure*}[t]
\begin{center}
\includegraphics[width=\linewidth]{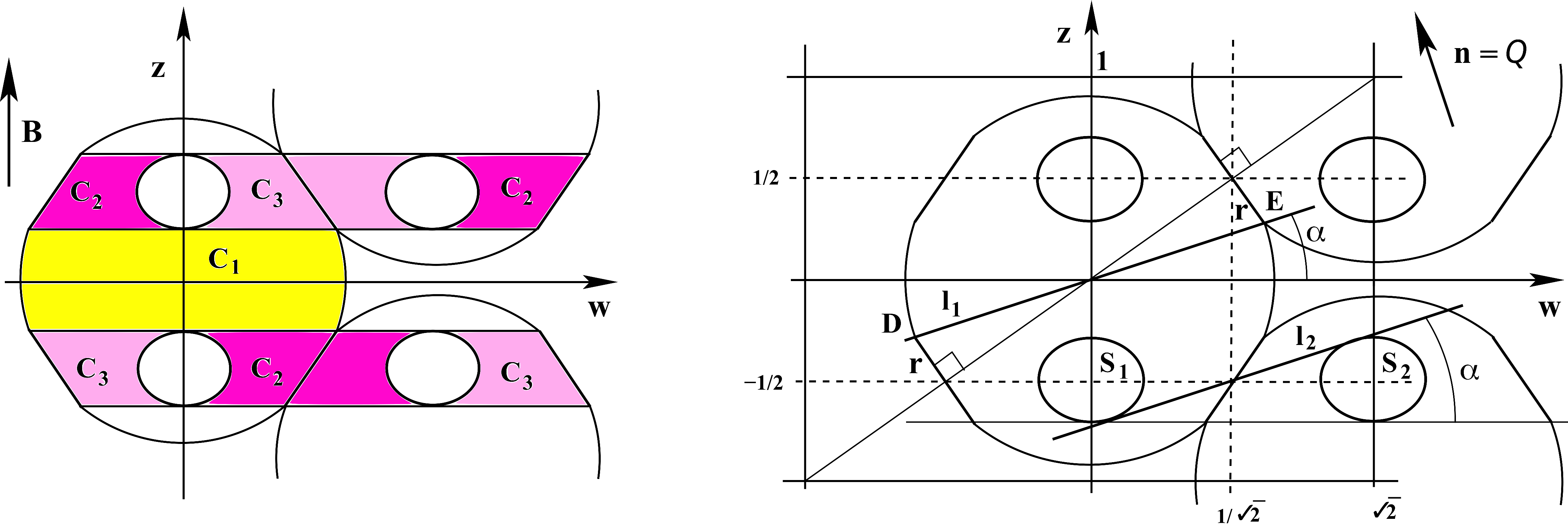}
\end{center}
\caption{Projection of the Fermi surface onto the plane 
$\, x + y = 0 \, $ for $\, {\bf B} \parallel z \, $ and 
disappearance of the cylinders $\, C_{1} \, $, $\, C_{2} \, $ 
and $\, C_{3} \, $ for $\, {\bf n} = Q \, $ (face-centered lattice).}
\label{Fig19}
\end{figure*}

 Let us now consider the direction $\, {\bf n} = Q \, $ 
(Fig. \ref{Fig8}). When projected onto the plane $\, x + y = 0 \, $, 
half of the circles $\, S_{i} \, $ are transformed into segments of 
length $ \, 2 r \, $, and the other half into ellipses with semi-axes 
$\, r \, $ and $\, r \sqrt{2/3} \, $ (Fig. \ref{Fig19}). The disappearance 
of the cylinders $\, C_{1} \, $, $\, C_{2} \, $ and $\, C_{3} \, $ is now 
determined by the condition that the line $\, {\bf l}_{1} \, $ passing 
through the points $\, D \, $ and $\, E \, $ and the line 
$\, {\bf l}_{2} \, $ tangent to the ellipses $\, S_{1} \, $ and 
$\, S_{2} \, $ (the projections of the planes 
$\, \Pi_{1,2} \perp {\bf B} \, $ onto the plane $\, x + y = 0 $) 
have the same inclination angle $\, \alpha \, $ (Fig. \ref{Fig19}). 
The angle $\, \alpha \, $ shown in Fig. \ref{Fig19} is the angle of 
inclination of $\, {\bf B} \, $ to the axis $\, z \, $ and determines 
the position of the point $\, Q \, $.

 As is easy to see, the equations of the ellipses $\, S_{1} \, $ 
and $\, S_{2} \, $ shown in Fig. \ref{Fig19} have the form
$$3 \left( z - 1/2 \right)^{2} \,\, + \,\, 
2 \, w^{2} \,\,\, = \,\,\, 2 \, r^{2} $$
and
$$3 \left( z - 1/2 \right)^{2} \,\, + \,\, 
2 \left( w - \sqrt{2} \right)^{2} \,\,\, = \,\,\, 2 \, r^{2} $$
 ($w = (y - x)/\sqrt{2}$), respectively.

 It can be seen that, on the one hand,
$$\tan \alpha \,\,\, = \,\,\, 
{1/2 \, - \, r \sqrt{2/3} \over 1/\sqrt{2} \, + \, r/\sqrt{3}} 
\,\,\, = \,\,\, 
{\sqrt{3} \, - \, 2 \sqrt{2} \, r \over \sqrt{6} \, + \, 2 \, r} $$
(disappearance of the cylinder $\, C_{1}$).

 On the other hand, the tangency condition of the line
$$\left( z - 1/2 \right) \,\,\, = \,\,\, \tan \alpha \, 
\left( w \, - \, 1/\sqrt{2} \right) $$
and the ellipses $\, S_{1} \, $ and $\, S_{2} \, $ leads to 
the relation
$$\tan \alpha \,\,\, = \,\,\, { 2 r \over \sqrt{3 - 6 r^{2}}} $$

 The given relations result in
$$r (Q) \,\,\, \simeq \,\,\, 0.264889 $$
$$\tan \alpha_{Q} \,\,\, \simeq \,\,\,  0.329890 $$
$$\widetilde{\epsilon}_{0} (Q) 
\,\,\, =  \,\,\, {3 \over 4} \,\, + r^{2} 
\,\,\, \simeq \,\,\,  0.820166 $$

 Thus, for the case of the face-centered lattice, we can write
$$\left[ \epsilon^{\cal B}_{1} , \, \epsilon^{\cal B}_{2} \right]
\,\,\, \simeq \,\,\, \left[ 0.817 , \, 0.820 \right] $$

\vspace{1mm}

 It can be seen that, according to our estimate, the interval 
$\, \left[ \epsilon^{\cal B}_{1} , \, \epsilon^{\cal B}_{2} \right] \, $ 
lies in the energy region where the genus of the Fermi surface is 4, 
and the increase in the genus of $\, S_{F} \, $ to 7 at 
$\, \epsilon_{F} = 1 \, $ does not affect the emergence of 
ultra-complex angular diagrams.

 Again abstracting from our normalization, we can say that, 
in the nearly free electron approximation for the face-centered 
cubic lattice, the interval of emergence of ultra-complex angular 
diagrams in the first conduction band lies at a distance of 
$\, \simeq \, 0.57 \, \epsilon_{\rm max} \, $ from its bottom 
and occupies $\, \simeq \, 0.2 \, \% \, $ of the band width.

\section{The body-centered cubic lattice}
\setcounter{equation}{0}

 In this section, we will consider the case of the body-centered cubic 
lattice. The reciprocal lattice in this case is face-centered cubic, 
and we will assume its edge length to be equal to 2. As in the previous 
case, we will be interested in the first conduction band in the nearly 
free electron approximation.

 The emergence of the first (compact) components of the Fermi surface 
now occurs near integer points $\, (m, n, l) \, $ of the 
$\, {\bf p}$ - space, such that
$$m + n + l \,\, = \,\, 2 s \,\,\, , \quad  s \in \mathbb{Z} $$

 As above, we will set
$$ \Delta \epsilon \left( {\bf p} \right) \,\, = \,\, 
\left( \Delta {\bf p} \right)^{2} $$
near these points.

\begin{figure*}[t]
\begin{center}
\includegraphics[width=0.9\linewidth]{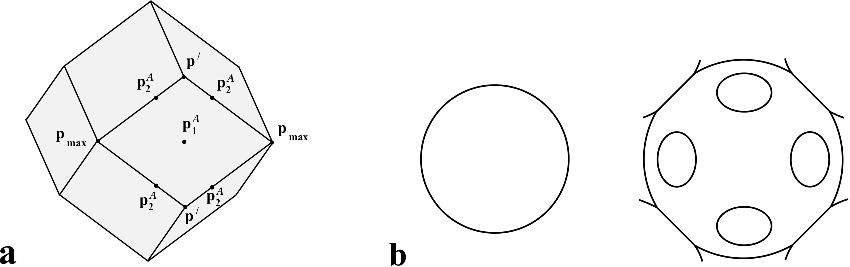}
\end{center}
\caption{(a) Brillouin zone for the body-centered cubic lattice 
(reciprocal lattice - face-centered cubic).
(b) Surgery of the Fermi surface upon passing through 
$\epsilon_{F} = 1/2$. }
\label{Fig20}
\end{figure*}

 The Brillouin zone now has the form shown in Fig. \ref{Fig20}, a.

 The Fermi surface remains spherical (in the torus $\mathbb{T}^{3}$) 
in the interval $\, \epsilon_{F} \, \in \, (0, 1/2) \, $.

 The value
$$ \epsilon_{F} \,\,\, = \,\,\, \epsilon^{\cal A}_{1} 
\,\,\, = \,\,\, 1/2 \,\,\, ,  
\quad \quad \left( p^{\cal A}_{1} \, = \, 1 / \sqrt{2} \right) 
\,\,\, , $$
corresponds to the first touch of the spheres in 
$\, {\bf p}$ - space (at the points $p^{\cal A}_{1}$), after which 
the Fermi surface acquires genus 6 (Fig. \ref{Fig20}, b).

 At the value
$$ \epsilon_{F} \,\,\, = \,\,\, \epsilon^{\cal A}_{2} 
\,\,\, = \,\,\, 2/3 \,\,\, ,  
\quad \quad \left( p^{\cal A}_{2} \, = \, \sqrt{2/3} \right) 
\,\,\, , $$
the spheres reach the edges of the Brillouin zone (at the points 
$p^{\cal A}_{2}$), and the Fermi surface again decomposes into 
components of genus 0, which vanish at the values
$$ \epsilon_{F} \,\,\, = \,\,\, \epsilon^{\prime}
\,\,\, = \,\,\, 3/4 \,\,\, ,  
\quad \quad \left( p^{\prime} \, = \, \sqrt{3}/2 \right) $$
and
$$ \epsilon_{F} \,\,\, = \,\,\, \epsilon_{\rm max}
\,\,\, = \,\,\, 1 \,\,\, ,  
\quad \quad \left( p_{\rm max} \, = \, 1 \right) $$

 We will, certainly, consider here the Fermi surfaces at
$\, \epsilon_{F} \in \left[ \epsilon^{\cal A}_{1} , \,
\epsilon^{\cal A}_{2} \right] \, $ 
(Fig. \ref{Fig20}, b).

 As in the previous sections, we will be interested here in 
the values of $\, \widetilde{\epsilon}_{0} ({\bf n}) \, $ at
the ``symmetric'' points of the boundary of $\, W_{1} \, $ 
(points $P$ and $Q$, Fig. \ref{Fig8}).

 To determine the structure of system (\ref{MFSyst}), 
as in the previous sections, we need to describe the cylinders 
of closed trajectories arising on the Fermi surface for 
$\, {\bf n} \in W_{1} \, $. Near the direction 
$\, {\bf n} \parallel z \, $, the number of such cylinders is 5, 
and they cut the Fermi surface into 4 tori with holes 
(carriers of open trajectories) $\, \mathbb{T}^{2}_{1,2,3,4} \, $, 
embedded in the three-dimensional torus
$$\mathbb{T}^{2}_{1,2,3,4} \,\,\, \subset \,\,\, 
\mathbb{T}^{2} \,\,\, = \,\,\, \mathbb{R}^{3} \Big/ L^{*} $$

\begin{figure}[t]
\begin{center}
\includegraphics[width=0.8\linewidth]{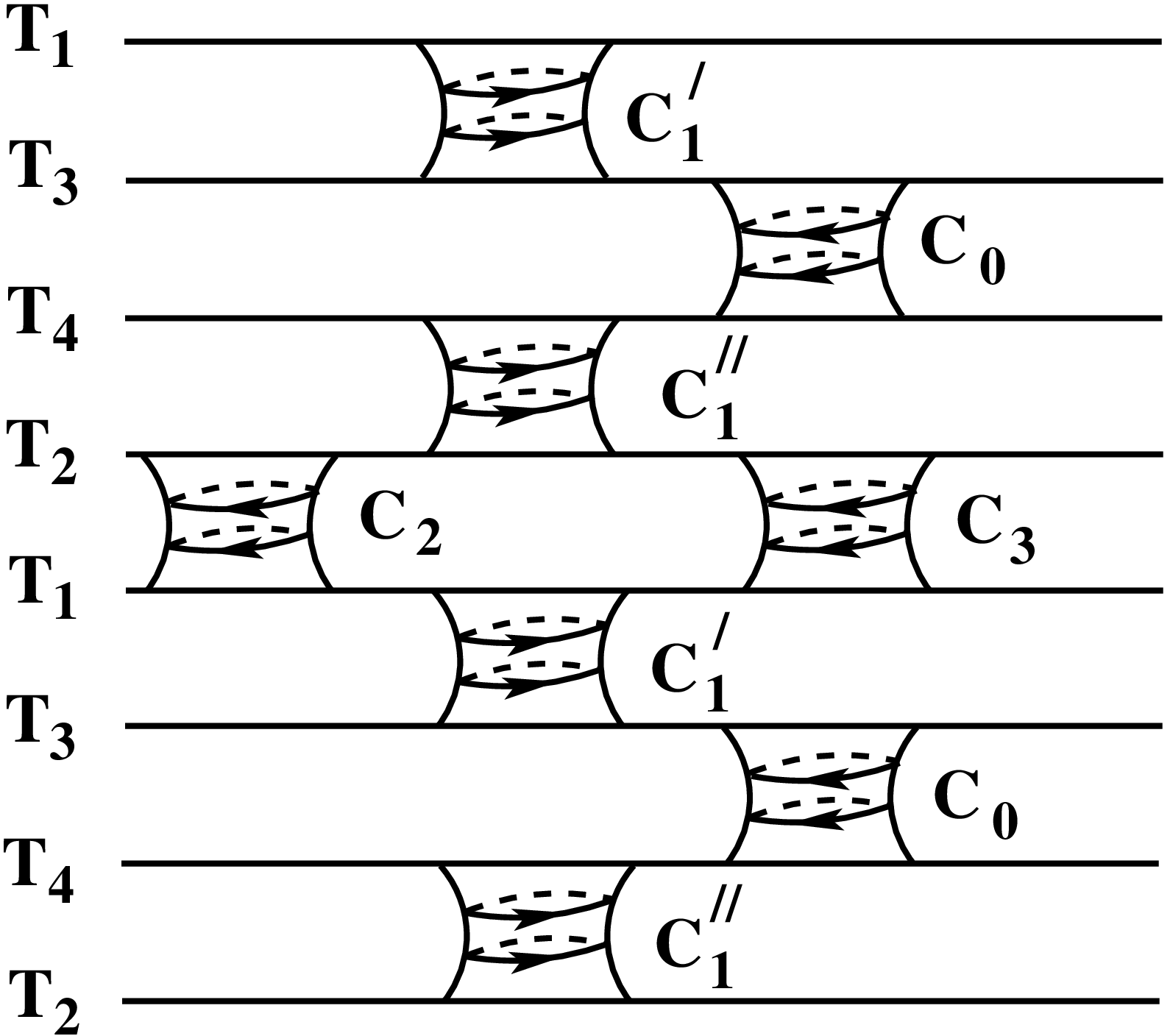}
\end{center}
\caption{``Scheme'' of the connection of carriers of open 
trajectories and cylinders of closed trajectories near the 
direction $\, {\bf B} \parallel z \, $ 
(body-centered cubic lattice).}
\label{Fig21}
\end{figure}

 In the full $\, {\bf p}$ - space, we then have 4 nonequivalent 
carriers of open trajectories, which are periodically deformed planes 
(with holes) parallel to the plane $\, x y \, $ (Fig. \ref{Fig21}).

 The height of the cylinders of closed trajectories is maximal 
at $\, {\bf B} \parallel z \, $, and their structure is in many 
ways analogous to the structure for the face-centered lattice. 
In particular, here also arise (hole-type) cylinders $\, C_{2} \, $ 
and $\, C_{3} \, $, completely analogous to the cylinders $\, C_{2,3} \, $ 
for the simple and face-centered lattices 
(see, e.g., Fig. \ref{Fig12}, \ref{Fig18}). As for the cylinder 
$\, C_{1} \, $ (Fig. \ref{Fig12}, \ref{Fig18}), now, due to the presence 
of additional ``handles'' (and a larger genus of the Fermi surface), 
it splits into two electron type cylinders $\, C_{1}^{\prime} \, $ and 
$\, C_{1}^{\prime\prime} \, $ and a hole type cylinder $\, C_{0} \, $ 
(similar to cylinders $\, C_{2} \, $ and $\, C_{3}$).

\begin{figure*}[t]
\begin{center}
\includegraphics[width=\linewidth]{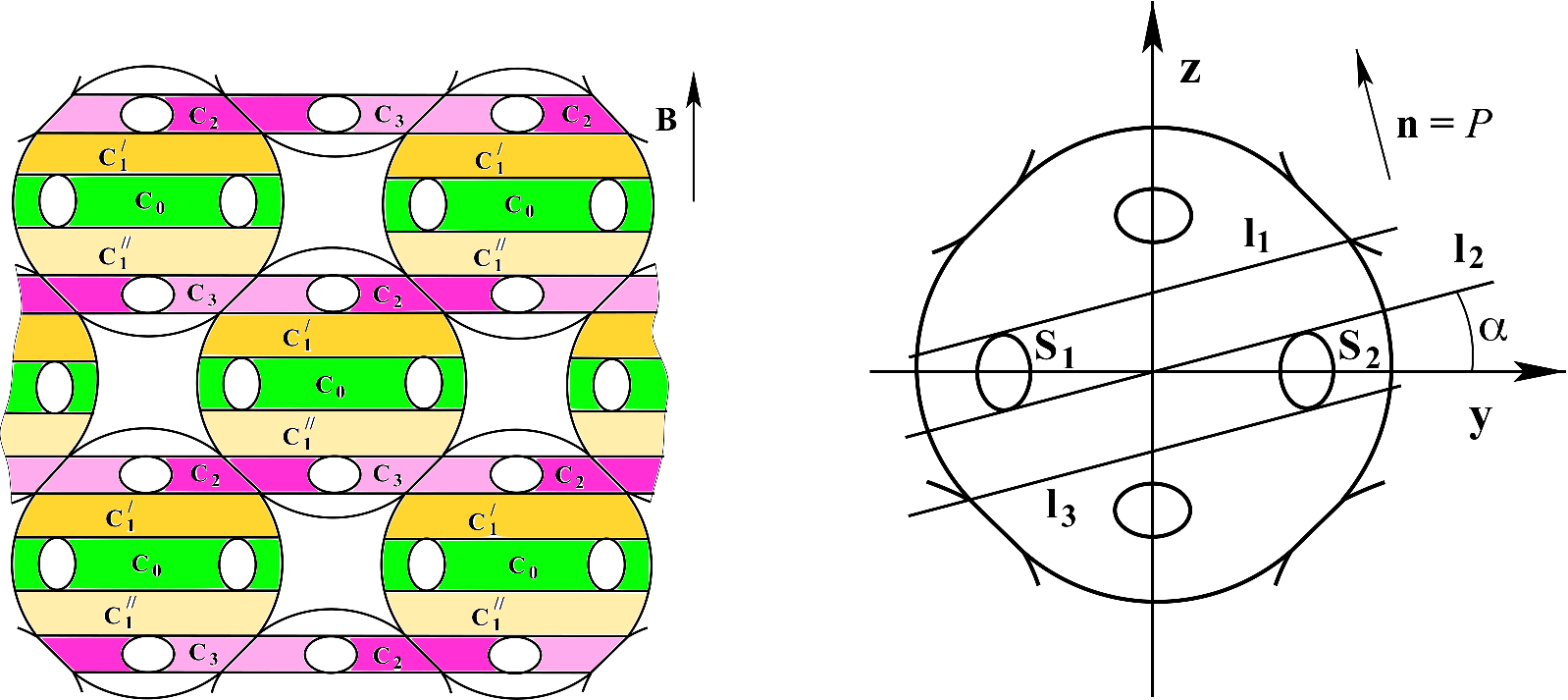}
\end{center}
\caption{Projections of cylinders of closed trajectories onto the 
plane $\, y z \, $ for $\, {\bf B} \parallel z \, $ and the 
disappearance of cylinders $\, C_{1}^{\prime} \, $, $\, C_{0} \, $ 
and $\, C_{1}^{\prime\prime} \, $ for $\, {\bf n} = P \, $ 
(body-centered lattice).}
\label{Fig22}
\end{figure*}

 Fig. \ref{Fig22} shows the arrangement of cylinders 
$\, C_{j} \, $ on the Fermi surface in $\, {\bf p}$ - space 
for $\, {\bf B} \parallel z \, $. Fig. \ref{Fig21} also shows 
the connection of the cylinders $\, C_{j} \, $ with 
the carriers of open trajectories $\, \mathbb{T}^{2}_{i} \, $ 
for directions of $\, {\bf B} \, $ close to the direction $\, z \, $.

 It is easy to see that, when $\, {\bf B} \, $ is deflected 
toward the point $\, P \, $, the cylinders $\, C_{2} \, $ and 
$\, C_{3} \, $ (they have a smaller height) disappear before 
the cylinder $\, C_{0} \, $. Thus, the point $\, P \, $ on the 
boundary of Zone $\, W_{1} \, $ is determined by the 
disappearance of the cylinders $\, C_{1}^{\prime} \, $ and 
$\, C_{1}^{\prime\prime} \, $ (of the electron type) and the 
cylinder $\, C_{0} \, $ (of the hole type).

 The disappearance of the cylinders $\, C_{1}^{\prime} \, $, 
$\, C_{0} \, $ and $\, C_{1}^{\prime\prime} \, $ at 
$\, {\bf n} = P \, $ is shown in Fig. \ref{Fig22}. The projections 
of the circles $\, S_{i} \, $ onto the plane $\, y z \, $ are now 
either segments of length $ \, 2 r \, $ or ellipses with semi-axes 
$\, r \, $ and $\, r / \sqrt{2} \, $. As is easy to see 
(Fig. \ref{Fig22}), the equations of the projections of
$\, S_{1} \, $ and $\, S_{2} \, $ have the form
$$2 \left( y \, + \, 1/2 \right)^{2} \,\, + \,\, z^{2} 
\,\,\, = \,\,\, r^{2} $$
and
$$2 \left( y \, - \, 1/2 \right)^{2} \,\, + \,\, z^{2} 
\,\,\, = \,\,\, r^{2} $$
respectively.

 The position of the point $\, P \, $ and the value $\, r (P) \, $ are 
determined by the equality of the inclination angles of the tangents 
(the projections of the planes $\, \Pi \perp {\bf B} \, $ onto the plane 
$\, y z $) to the projections of the circles $\, S_{i} \, $ onto the plane 
$\, y z \, $, as shown in Fig. \ref{Fig22}. As can be verified, 
the condition of tangency of the line $\, {\bf l}_{2} \, $ with the 
ellipses $\, S_{1} \, $ and $\, S_{2} \, $ has the form
$$\tan \alpha \,\,\, = \,\,\, {2 r \over \sqrt{1 - 2 r^{2}}} $$

 The line $\, {\bf l}_{1} \, $ is given by the equation
$$z \,\,\, = \,\,\, {2 r \over \sqrt{1 - 2 r^{2}}}
\left( y + 1 \right) $$

 According to Fig. \ref{Fig22}, it must contain the point
$$\left( {1 \over 2} + {r \over \sqrt{2}} , \, 
{1 \over 2} - {r \over \sqrt{2}} \right) \,\,\, , $$
which implies
$$\left( 1 - r \sqrt{2} \right) \sqrt{1 - 2 r^{2}} \,\,\, = \,\,\,
2 r \left( 3 + r \sqrt{2} \right) $$

 From the above relations we directly obtain
$$r (P) \,\,\, \simeq \,\,\, 0.126932 $$
$$\tan \alpha_{P} \,\,\, \simeq \,\,\, 0.258056 $$
$$\widetilde{\epsilon}_{0} (P) \,\,\, \simeq \,\,\, 0.516112 $$

\vspace{2mm}

\begin{figure*}[t]
\begin{center}
\includegraphics[width=\linewidth]{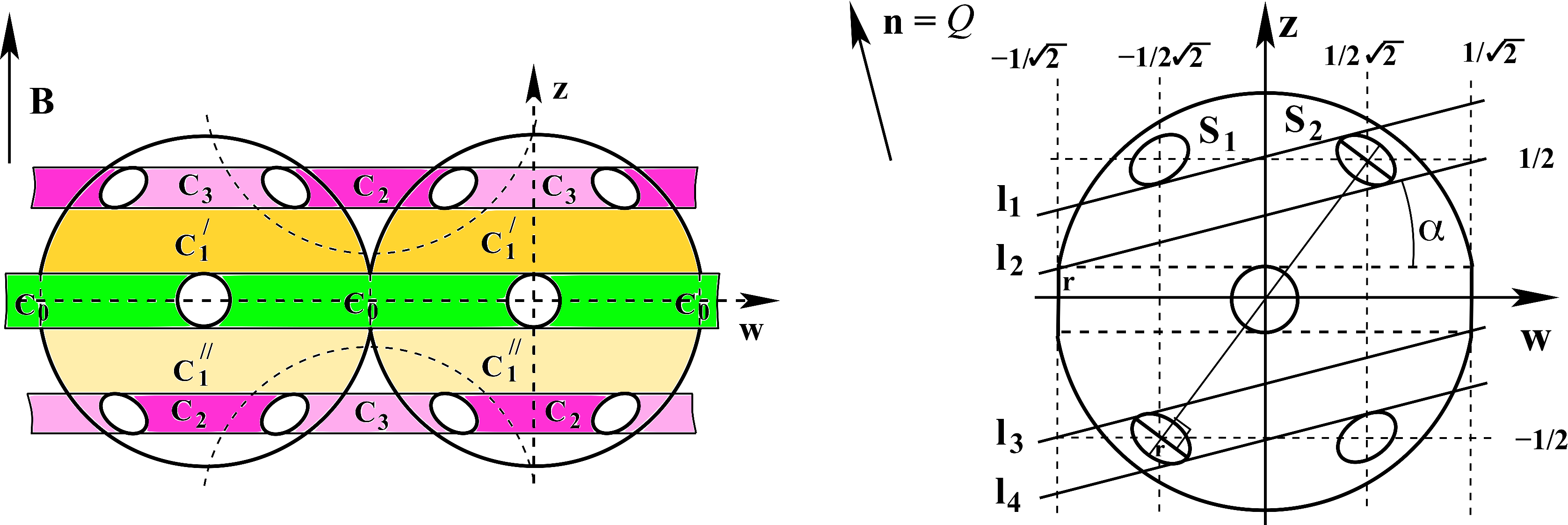}
\end{center}
\caption{Projections of cylinders of closed trajectories onto the 
plane $\, x + y = 0 \, $ for $\, {\bf B} \parallel z \, $ and the 
disappearance of cylinders $\, C_{1}^{\prime} \, $, 
$\, C_{1}^{\prime\prime} \, $, $\, C_{2} \, $ and $\, C_{3} \, $ 
for $\, {\bf n} = Q \, $ (body-centered lattice).}
\label{Fig23}
\end{figure*}

 Let us now consider the direction $\, {\bf n} = Q \, $. 
Fig. \ref{Fig23} shows the projection of the Fermi surface 
onto the plane $\, x + y = 0 \, $. The projections of the 
circles $\, S_{i} \, $ represent here either segments of 
length $\, 2 r \, $, or circles of radius $\, r \, $, or 
ellipses with semi-axes $\, r \, $ and $\, r/2 \, $. 
As is easy to see, the centers of the ellipses $\, S_{1} \, $ 
and $\, S_{2} \, $ lie at the points with coordinates
$$w \,\,\, = \,\,\, - {1 \over 2 \sqrt{2}} \quad , \quad \quad
z \,\,\, = \,\,\, {1 \over 2} $$
$$w \,\,\, = \,\,\, {1 \over 2 \sqrt{2}} \quad , \quad \quad
z \,\,\, = \,\,\, {1 \over 2} $$
($w \, = \, (y - x)/\sqrt{2}$).

 As is not also difficult to show, the major semi-axes of the 
ellipses $\, S_{1} \, $ and $\, S_{2} \, $ are orthogonal to the 
segments connecting their centers with the origin.

 Now it can be seen that, when $\, {\bf B} \, $ is deflected 
towards the point $\, Q \, $, the cylinders $\, C_{0} \, $ 
(their projections onto the plane $\, x + y = 0 \, $ have a much 
greater length) disappear before the cylinders $\, C_{2} \, $ and 
$\, C_{3} \, $. Thus, the point $\, Q \, $ on the boundary of 
Zone $\, W_{1} \, $ is determined by the disappearance of the 
cylinders $\, C_{1}^{\prime} \, $ and $\, C_{1}^{\prime\prime} \, $ 
(of the electron type) and the cylinders $\, C_{2} \, $ and 
$\, C_{3} \, $ (of the hole type).

 As before, the projections of planes orthogonal to 
$\, {\bf B} \, $ onto the plane $\, x + y = 0 \, $ here represent 
lines with an inclination angle $\, \alpha \, $ with respect to 
the axis $\, w \, $. The condition for the simultaneous 
disappearance of the cylinders $\, C_{1}^{\prime} \, $, 
$\, C_{1}^{\prime\prime} \, $, $\, C_{2} \, $ and $\, C_{3} \, $ 
is determined by the simultaneous tangency of the projections 
$\, S_{i} \, $ by such lines, as shown in Fig. \ref{Fig23}.

 Ellipses $\, S_{1} \, $ and $\, S_{2} \, $ are given by 
the equations
$$4 \left( \sqrt{2} z - w - {3 \over 2 \sqrt{2}} \right)^{2} 
\,\, + \,\, \left( z + \sqrt{2} w  \right)^{2}
\,\,\, = \,\,\, 3 r^{2} $$
and
$$4 \left( \sqrt{2} z + w - {3 \over 2 \sqrt{2}} \right)^{2} 
\,\, + \,\, \left( z - \sqrt{2} w  \right)^{2}
\,\,\, = \,\,\, 3 r^{2} $$
respectively.

 The equations of the lines $\, {\bf l}_{1} \, $ and 
$\, {\bf l}_{2} \, $, as is not difficult to show, 
have the form
$$z \,\,\, = \,\,\, w \, \tan \alpha  \,\, + \,\, 
1 \, - \, \sqrt{2} \, \tan \alpha \, - \, r $$
and
$$z \, - \, r  \,\,\, = \,\,\,  
\left( w + {1 \over \sqrt{2}} \right) \tan \alpha $$

 As can be shown, the conditions for the correct tangency of the 
lines $\, {\bf l}_{1} \, $, $\, {\bf l}_{2} \, $ and the ellipses 
$\, S_{1} \, $, $\, S_{2} \, $ are given by the relations
\begin{multline*}
\sqrt{2} \, r \,\, \sqrt{ \left( \sqrt{2} + \tan \alpha \right)^{2} 
\, + \, 4 \left( 1 - \sqrt{2} \tan \alpha \right)^{2}} 
\,\,\, =  \\
= \,\,\, 5 \sqrt{3} \, \tan \alpha \,\, + \,\, 
2 \sqrt{6} \, r \,\, - \,\, \sqrt{6} 
\end{multline*}
and
\begin{multline*}
\sqrt{2} \, r \,\, \sqrt{ \left( \sqrt{2} - \tan \alpha \right)^{2} 
\, + \, 4 \left( 1 + \sqrt{2} \tan \alpha \right)^{2}} 
\,\,\, =  \\
= \,\,\, \sqrt{6} \,\, - \,\, 3 \sqrt{3} \, \tan \alpha 
\,\, - \,\, 2 \sqrt{6} \, r  \,\,\, , 
\end{multline*}
which implies
$$r (Q) \,\,\, \simeq \,\,\, 0.123591 $$
$$\tan \alpha_{Q} \,\,\, = \,\,\, 0.255360 $$
$$\widetilde{\epsilon}_{0} (Q) 
\,\,\, \simeq \,\,\, 0.515275 $$

\vspace{1mm}

 Thus, for the case of the body-centered lattice, we can write
$$\left[ \epsilon^{\cal B}_{1} , \, \epsilon^{\cal B}_{2} \right]
\,\,\, \simeq \,\,\, \left[ 0.5153 , \, 0.5161 \right] $$

\vspace{1mm}

 Abstracting from our normalization, we can say that, in the 
nearly free electron approximation for the body-centered cubic lattice, 
the interval of emergence of ultra-complex angular diagrams in the 
first conduction band lies at a distance of 
$\, \simeq \, 0.515 \, \epsilon_{\rm max} \, $ from its bottom 
and occupies $\, \simeq \, 0.1 \, \% \, $ of the band width.

\section{Conclusion}
\setcounter{equation}{0}

 We estimate the ranges of occurrence of ultra-complex conductivity 
diagrams in the nearly free electron approximation for metals with 
cubic symmetry. The results of the study show that the occurrence of 
such diagrams corresponds to extremely narrow energy intervals 
$\, \epsilon_{F} \in \left[ \epsilon^{\cal B}_{1} , \, 
\epsilon^{\cal B}_{2} \right] \, $ inside the conduction band, which 
coincides with similar estimates in the tight binding approximation 
(\cite{TightBind}). In our opinion, the obtained results are largely 
explained by the high symmetry of the crystal lattices, as well as 
the simplest analytical properties of the relations 
$\, \epsilon ({\bf p}) \, $ considered in these studies. In particular, 
we expect wider ranges of occurrence of ultracomplex angular conductivity 
diagrams for conductors with more complex relations 
$\, \epsilon ({\bf p}) \, $. In general, to observe ultra-complex 
angular conductivity diagrams, external influences on the crystal, 
shifting the position of $\, \epsilon_{F} \, $ within the conductivity 
band, are probably necessary.

\end{document}